\documentclass[fleqn,usenatbib]{mnras}

\usepackage{newtxtext,newtxmath}

\usepackage[T1]{fontenc}

\DeclareRobustCommand{\VAN}[3]{#2}
\let\VANthebibliography\thebibliography
\def\thebibliography{\DeclareRobustCommand{\VAN}[3]{##3}\VANthebibliography}

\usepackage{graphicx}	

\usepackage{amsmath}	
\usepackage{amssymb}	
\usepackage{multirow}
\usepackage{booktabs}
\usepackage{caption,subcaption}
\usepackage{afterpage}
\usepackage{siunitx}

\usepackage[normalem]{ulem}

\title[Y-dwarf near-infrared photometry with HST]{The Y Dwarf Population with HST: unlocking the secrets of our coolest neighbours -- III. Near-Infrared Photometry \thanks{
Based on observations with the NASA/ESA {\it Hubble
Space Telescope}, obtained at the Space Telescope Science Institute,
which is operated by AURA, Inc., under NASA contract NAS 5-26555.
}
}

\author[C. Fontanive et al.]{
Clémence Fontanive,$^{1,2}$\thanks{E-mail: \href{mailto:clemence.fontanive@roe.ac.uk} {clemence.fontanive@roe.ac.uk}}
Luigi R. Bedin,$^{3}$
Mark W. Phillips,$^{1}$
Michele Scalco,${^4}$
Loïc Albert,$^{2}$
\newauthor
Daniella C. Bardalez Gagliuffi,$^{5}$
and Beth Biller$^{1}$
\\
$^{1}$SUPA, Institute for Astronomy, University of Edinburgh, Blackford Hill, Edinburgh EH9 3HJ, UK\\
$^{2}$Trottier Institute for Research on Exoplanets, Université de Montréal, Montréal H3C 3J7, Québec, Canada\\
$^{3}$INAF-Osservatorio Astronomico di Padova, Vicolo dell’Osservatorio 5, I-35122 Padova, Italy\\
$^{4}$Department of Astronomy, Indiana University, Swain West, 727 E. 3rd Street, Bloomington, IN 47405, USA\\
$^{5}$Department of Physics \& Astronomy, Amherst College, 25 East Drive, Amherst, MA 01002, USA\\
}

\date{Accepted 2026 February 9. Received 2026 January 29; in original form 2025 December 3}

\pubyear{2026}

\defcitealias{Fontanive2021}{Paper~I}
\defcitealias{Fontanive2025}{Paper~II}

\begin{document}
\label{firstpage}
\pagerange{\pageref{firstpage}--\pageref{lastpage}}
\maketitle

\begin{abstract}
Y dwarfs represent the coldest class of brown dwarfs, with effective temperatures below 500\,K, and provide unique analogues to cold giant exoplanets. We present a large compilation of uniform near‑infrared photometry from the \textit{Hubble Space Telescope} for 21 Y dwarfs across multiple WFC3/IR filters, including the F105W, F125W and F160W bands. We employed refined PSF‑fitting and calibration procedures to reach photometric uncertainties at the 0.02--0.05\,mag level for most targets. Combined with precise parallax measurements, our data reveal well-defined Y-dwarf sequences in near-infrared colour–magnitude diagrams, observed with a markedly improved tightness. Known photometric trends emerge with minimal scatter, including the continuous redward progression in F125W--F160W with decreasing temperature, and the blueward trend in F105W--F125W with possible hints of a reversal around 350\,K. Comparisons to the \texttt{ATMO}, \texttt{Sonora\,Elf\,Owl}, and \texttt{Lacy\,\&\,Burrows} atmospheric models highlight systematic discrepancies, in particular F105W--F125W and F105W--F160W colours predicted to be too red. Low‑metallicity grids provide the best fits to the global Y-dwarf population, although closer inspection across wavelengths shows that these preferences likely reflect compensating effects in missing or incomplete physics rather than true population‑level abundances. While some atmospheric diversity is expected among Y dwarfs, their tight observational sequences and systematic offsets from model predictions reveal that key physical and chemical processes remain inadequately captured in current grids. Our results underscore the importance of high‑precision, internally consistent datasets in robustly tracing the Y-dwarf cooling sequence and providing the empirical constraints needed to advance theoretical models at the lowest temperatures.
\end{abstract}




\section{Introduction}

In contrast to stars, substellar objects lack hydrogen fusion, causing giant planets and brown dwarfs to continuously cool and fade over time. At the cold and faint end of this evolution, the Y spectral sequence \citep{Cushing2011} classifies the coldest extra-solar worlds, with effective temperatures T$_\mathrm{eff} < 500$\,K. With no host star flooding their light, free-floating Y dwarfs are amenable to direct atmospheric characterisation, offering excellent proxies to investigate in detail cold planetary atmospheres \citep{Beichman2014,Skemer2016,Leggett2019b}. In particular, the most extreme members of the Y class known to date ($\sim$250--350\,K; \citealp{Luhman2014,Marocco2019,BardalezGagliuffi2020,Meisner2020,Calissendorff2023}) are only a few hundred K warmer than Jupiter (130\,K) and provide valuable analogues to directly-imaged exoplanets like Eps\,Ind\,Ab (275\,K; \citealp{Matthews2024}), 14\,Her\,c (300\,K; \citealp{BardalezGagliuffi2025}) and TWA\,7b (315\,K; \citealp{Lagrange2025}), now within the reach of the \textit{James Webb Space Telescope} (\textit{JWST}). However, despite their significance, the intrinsic faintness of Y dwarfs has long rendered them accessible to only a handful of instruments, and mostly kept them out of reach for spectroscopic observations, severely hindering our understanding of the structure and composition of their cold atmospheres. 

Until recently, the lack of reliable distance estimates was a major obstacle to our understanding of the origins and spectro-photometric characteristics of Y dwarfs. Parallactic distance measurements are essential for calibrating their absolute properties, yet their extreme faintness makes the derivation of astrometric solutions particularly challenging \citep{Tinney2014,Beichman2014}. The \textit{Hubble Space Telescope} (\textit{HST}) has played a key role in overcoming this limitation, thanks to an extensive legacy of Y-dwarf observations acquired over the last 10--15\,years that now enable precise astrometry for a growing sample of Y dwarfs. Leveraging \textit{HST}'s deep sensitivity and excellent stability achieved from space, our team recently delivered high-precision parallax measurements for 15 Y dwarfs in \citet{Fontanive2021,Fontanive2025} (hereafter \citetalias{Fontanive2021} and \citetalias{Fontanive2025}, respectively). Complementary long-term efforts with the \textit{Spitzer} telescope further expand the catalogues of distance measurements now available for the nearby Y-dwarf population \citep{Martin2018,Kirkpatrick2019,Kirkpatrick2021,Marocco2019,Marocco2021}, which is finally amenable to robust calibrations levels.

With the knowledge of precise distances, absolute magnitudes may be plotted against photometric colours via colour–magnitude diagrams (CMDs), widely used in astronomy to trace the evolution and fundamental properties of stellar objects. For brown dwarfs, CMDs represent extremely powerful tools to compare the photometric properties of individual objects to well-calibrated standards, enabling probes of secondary attributes through the identification of outliers to the standardised locus. For example, overluminous sources may be indicative of unresolved binarity \citep{Manjavacas2013,Tinney2014}. Likewise, excessively red or blue colours can trace a deviant composition or youth, or be evidence for diverse atmospheric features like clouds or variability \citep{Knapp2004,Cruz2007,Cruz2009,Kirkpatrick2021b}. At the population level, CMDs have proven incredibly useful in brown dwarf science, highlighting key trends like the strong reddening of L dwarfs \citep{Marley2002} or the abrupt L/T transition \citep{DupuyLiu2012}. CMD analyses hence help constrain the evolution in photometric properties as a function of temperature, and have greatly enhanced our understanding of the complex physics and chemistry governing substellar atmospheres.

In Y dwarfs, information from observed CMDs has been central for providing first‑order estimates of spectral types and effective temperatures (e.g., \citealp{Luhman2012,Leggett2013,Leggett2016,Martin2018,Marocco2019,BardalezGagliuffi2020,Leggett2021}). Empirical relationships between colours and absolute magnitudes are also often used for distance determinations of new candidates (e.g., \citealp{Kirkpatrick2019,Meisner2020,Kirkpatrick2021}). Departures from local sequences have further served as diagnostics of surface gravity, metallicity, cloud properties, and disequilibrium chemistry \citep{Faherty2014,Luhman2014,Leggett2013,Leggett2017,Kirkpatrick2021b}, although such interpretations remain model‑dependent. On the modelling side, CMDs continue to play a crucial role in the development and testing of atmospheric models for Y dwarfs. Comparisons between theoretical predictions and observed sequences remain one of the primary diagnostics for evaluating the impact of different physical and atmospheric assumptions, and for identifying shortcomings in current models towards reproducing the bulk properties of the observed population (e.g., \citealp{Phillips2020,LacyBurrows2023}), for which constraining observational sequences with high levels of precision and accuracy is key.

Y dwarfs emit most of their flux at mid-infrared (MIR; $>$2.5\,$\mu$m) wavelengths due to their extremely cold temperatures. Facilities probing this spectral range (\textit{WISE}, \textit{Spitzer}) are hence responsible for the discovery of almost all known ultracool brown dwarfs, and photometry at 3.6\,$\mu$m and 4.5\,$\mu$m is available for the full set of targets populating the Y sequence \citep{Kirkpatrick2021}. 
Near-infrared (NIR; 0.8--2.5\,$\mu$m) observations are however challenging for the very faintest objects, that are too dim for surveys like \textit{2MASS} and rarely reach accuracies better than $\sim$0.1--0.2\,mag in ground-based photometry. As a result, the NIR data currently available for the Y-dwarf population originate from various facilities and instruments, each with their own photometric system, and remain somewhat inhomogeneous in quantity, quality and spectral coverage (e.g., \citealp{Leggett2017,Leggett2021}). Although growing sets of \textit{JWST} photometry for Y dwarfs are slowly becoming available both from imaging observations \citep{Albert2025} and synthetic photometry from spectra \citep{Beiler2024,Leggett2025}, this heterogeneity still limits our ability to trace population-level atmospheric properties at the depths probed by NIR wavelengths down to the coldest temperatures.

In this paper, we present a homogeneous set of high-precision multi-band NIR photometry from \textit{HST} for 21 Y dwarfs, enabling investigations of well calibrated and internally consistent colour–magnitude diagrams for a large fraction of the current Y-dwarf population. We present the observations in Section~\ref{observations}, along with the data reduction in Section~\ref{reduction}, and the photometric calibration and flux measurements in Section~\ref{photometry}, respectively. Analyses of the population-level trends in Y-dwarf brightness and colours are presented in Sections~\ref{results}, together with detailed comparisons to theoretical models in Section~\ref{models}. We discuss our results in Section~\ref{discussion} and summarise our conclusions in Section~\ref{conclusions}.

\section{Summary of Observations}
\label{observations}

\begin{table*}
    \caption{\textit{HST} WFC3/IR observations used for the photometry presented in this work for the 6 targets that were not part of the astrometric work from \citetalias{Fontanive2025}.}
\footnotesize
\resizebox{\textwidth}{!}{
    \centering
    \begin{tabular}{l l l l l l l l l}
\hline\hline
Target Name & Short Name & SpT & Disc.\ \& SpT Ref. & Program ID & Filter & N$_\mathrm{im}$ & t$_\mathrm{exp}$ (s) & Obs. Date \\
\hline \\[-0.2cm]
WISE J033605.05$-$014350.4$^\mathrm{a}$  & WISE\,0336$-$0143AB & Y0 & (a), (b) & GO-17466 & F105W, F160W & 2, 6 & 556, 1643 & 2023-12-12 \\
                            &                  &    &                           & GO-16229 & F105W, F160W & 2, 6 & 556, 1643 & 2021-08-06 \\ 
                            &                  &    &                           & GO-16229 & F105W, F160W & 2, 6 & 556, 1643 & 2021-01-30 \\
                            &                  &    &                           & GO-12970 & F125W & 4 & 2412 & 2013-02-04 \\[0.1cm]
WISE J035934.06$-$540154.6  & WISE\,0359$-$5401 & Y0 & (c), (c)    & GO-16229 & F160W & 8 & 2399 & 2021-08-14 \\
                            &                  &    &                           & GO-16229 & F105W, F125W & 4, 4 & 1212, 1187 & 2021-04-28 \\
                            &                  &    &                           & GO-15201 & F127M & 4 & 1212 & 2019-01-27 \\
                            &                  &    &                           & GO-12330 & F140W & 4 & 312 & 2011-08-09 \\[0.1cm]
WD 0806$-$661B & WD 0806$-$661B & Y1 & (d), (e)  & GO-13428 & F105W & 9 & 9027 & 2014-02-25 \\
                            &                  &    &                           & GO-13428 & F160W & 6 & 6018 & 2014-02-22 \\
                            &                  &    &                           & GO-13428 & F125W & 3 & 3009 & 2014-02-15 \\
                            &                  &    &                           & GO-13428 & F127M & 3 & 3009 & 2014-02-11 \\
                            &                  &    &                           & GO-12815 & F110W & 18 & 18052 & 2013-02-08 \\[0.1cm]
WISE J085510.83$-$071442.5  & WISE\,0855$-$0714 & Y4 & (f), (e)    & GO-17403 & F160W & 10 & 3992 & 2024-11-07 \\
                            &                  &    &                           & GO-17080 & F160W & 10 & 4542 & 2023-05-11 \\
                            &                  &    &                           & GO-17080 & F160W & 10 & 4542 & 2022-11-07 \\
                            &                  &    &                           & GO-14157 & F127M & 6 & 5418 & 2016-04-06 \\
                            &                  &    &                           & GO-14157 & F105W, F127M & 6, 12 & 5418, 10836 & 2016-03-27/28 \\
                            &                  &    &                           & GO-14233 & F160W & 48 & 15544 & 2016-03-27 \\
                            &                  &    &                           & GO-14157 & F105W & 12 & 10836 & 2016-03-22/23 \\
                            &                  &    &                           & GO-14233 & F160W & 32 & 10396 & 2016-03-15 \\
                            &                  &    &                           & GO-14233 & F105W, F125W & 12, 12 & 3836, 3636 & 2016-03-01 \\
                            &                  &    &                           & GO-13802 & F110W & 6 & 5418 & 2015-04-11 \\
                            &                  &    &                           & GO-13802 & F110W & 6 & 5418 & 2015-03-03 \\
                            &                  &    &                           & GO-13802 & F110W & 6 & 5418 & 2014-11-25 \\[0.1cm]
WISEP J140518.40$+$553421.5 & WISE\,1405$+$5534 & Y0.5p & (g), (h) & GO-16229 & F125W, F160W & 2, 6 & 606, 1818 & 2020-12-03 \\
                            &                  &    &                           & GO-12972 & F110W, F160W & 4, 5 & 1212, 1515 & 2013-04-23 \\
                            &                  &    &                           & GO-12970 & F105W & 4 & 412 & 2013-04-18 \\
                            &                  &    &                           & GO-12330 & F140W & 4 & 412 & 2011-03-14 \\[0.1cm]
WISEA J235402.79$+$024014.1 & WISE\,2354$+$0240 & Y1 & (h), (h)    & GO-16229 & F105W, F160W & 2, 6 & 556, 1643 & 2020-12-13 \\ 
                            &                  &    &                           & GO-15201 & F127M & 4 & 1197 & 2018-06-18 \\
                            &                  &    &                           & GO-13178 & F125W & 8 & 824 & 2013-09-22 \\
                            &                  &    &                           & GO-13178 & F125W & 4 & 2412 & 2013-05-27 \\
\hline\\ [-2.5ex]
    \multicolumn{9}{l}{
    \begin{minipage}{\linewidth}
        \textbf{Notes.}\\
        $^\mathrm{a}$ Found with \textit{JWST}/NIRCam to be a 1-au binary by \citet{Calissendorff2023}, unresolved in \textit{HST} observations.\\
        Discovery and spectral type references: (a) \citet{Mace2013}; (b) \citet{Martin2018}; (c) \citet{Kirkpatrick2012}; (d) \citet{Luhman2011}; (e) \citet{Kirkpatrick2021}; (f) \citet{Luhman2014}; (g) \citet{Cushing2011}; (h) \citet{Schneider2015}.
    \end{minipage}}
\end{tabular}
}
\label{t:observations}
\end{table*}

This study builds on \textit{HST} program GO-16229 (PI Fontanive), which was designed to obtain precise astrometry and multi-band photometry for a well-defined sample of Y dwarfs. The program targeted 19 nearby, confirmed free-floating Y dwarfs that had at least one prior \textit{HST} imaging epoch, allowing for long-baseline astrometric solutions to be derived from combinations of archival and newly acquired data. Together with WISE\,0855$-$0714, the coldest member of the Y sequence known to date \citep{Luhman2014}, and WD\,0806$-$661B, a wide-orbit companion to a nearby white dwarf \citep{Luhman2011}, which already possessed the full suite of archival observations required to meet the science goals of this campaign, our sample comprises a total of 21 Y-dwarf targets.

The GO-16229 observations consisted of carefully timed WFC3/IR imaging, scheduled to ensure optimal baselines and annual parallax coverage for astrometric solutions. All observations were obtained in the F105W, F125W, and F160W filters, which cover the $Y$-, $J$- and $H$-band emission peaks in brown dwarf spectra in the near-infrared range. The program was designed such that each target had at least one measurement in each of these bands when combining our new observations and existing archival data. Observations were conducted using optimised dithering strategies to mitigate detector systematics, improve spatial resolution, and facilitate the removal of cosmic rays and bad pixels (see \citetalias{Fontanive2021} and \citetalias{Fontanive2025} for details). Exposure times varied across filters and epochs, with deeper integrations in F160W to maximise signal-to-noise for our extremely faint targets. 

Astrometric results for 15 of the 19 Y dwarfs from GO-16229 were presented in \citetalias{Fontanive2025}. From the remaining 4 targets, 3 were excluded from the astrometric work because too few reference \textit{Gaia} stars were available in the fields of view for our current methodologies to be successfully implemented (WISE\,0359$-$5401, WISE\,1206$+$8401 and WISE\,2354$+$0240). The fourth target (WISE\,0336$-$0143) was left out of \citetalias{Fontanive2025} because it was since discovered with \textit{JWST}/NIRCam to be a binary system unresolvable in existing \textit{HST} data \citep{Calissendorff2023}, and more advanced astrometric analyses accounting for the binary nature of the system await completion of ongoing follow-up programs (\textit{HST} GO-17466 \& GO-17841, PI Fontanive).

In this paper, we present complete photometry sets in the WFC3/IR F105W, F125W and F160W filters for all 19 targets from the original GO-16229 sample, including the 4 sources left out of \citetalias{Fontanive2025}, to which we add WISE\,0855$-$0714 and WD\,0806$-$661B. We incorporate all available \textit{HST} data from our own campaign and any earlier WFC3 programs aimed at these objects: GO-12330 (PI Kirkpatrick; \citealp{Kirkpatrick2012}), GO-12544 (PI Cushing; \citealp{Beichman2014}) GO-12815 (PI Luhman; \citealp{Luhman2014b}), GO-12873 (PI Biller; \citealp{Fontanive2018}), GO-12970 (PI Cushing; \citealp{Schneider2015}), GO-12972 (PI Gelino), GO-13178 (PI Kirkpatrick; \citealp{Schneider2015}), GO-13428 (PI Gelino; \citealp{Leggett2017}), GO-13802 (PI Luhman; \citealp{LuhmanEsplin2016}), GO-14157 (PI Luhman; \citealp{LuhmanEsplin2016}), GO-14233 (PI Schneider; \citealp{Schneider2016}), GO-15201 (PI Fontanive; \citealp{Fontanive2023}), GO-17080 (PI Bedin), GO-17403 (PI Bedin), GO-17466 (PI Fontanive). From these, we obtain NIR flux measurements in the F105W, F125W and F160W bands for all 21 Y dwarfs, in addition to the F127M or F140W filters for about half of the targets, and F110W for 3 objects.
Details from all considered observations for the 15 targets from \citetalias{Fontanive2025} are presented in Table~1 of that paper. In Table~\ref{t:observations} of the present paper, we summarise in the same format the datasets used for the additional 4 targets from the GO-16229 sample, along with WISE\,0855$-$0714 and WD\,0806$-$661B.

\section{Data Reduction}
\label{reduction}

\subsection{First-Pass Photometry}

Data reduction was carried out following the methodology described in \citetalias{Fontanive2021} and \citetalias{Fontanive2025} (see also \citealt{BedinFontanive2018,BedinFontanive2020}). Below, we provide a summary of these procedures, while referring readers to the original works for a more comprehensive description.

We started by extracting the positions and magnitudes of all sources detected in each WFC3/IR bias-, flat-field-, dark-, and CTE- (charge transfer efficiency) corrected image (\texttt{\_flt}) using the publicly available software developed by \citet{AndersonKing2006}, in its most updated version \texttt{hst1pass} \citep{Anderson2022}.
Following the procedure in \citet{Anderson2016}, we started from \textit{effective} Point Spread Function ($e$PSF) libraries, and refined these $e$PSFs by iterating between the library models and the residuals of the brightest stars in the field of view, to determine the best spatially variable $e$PSF for each image. Stellar positions and fluxes were then derived by fitting each source with the optimised $e$PSF, according to the perturbation of the library $e$PSF as described by \citet{AndersonKing2006}. 
Stellar positions in each single-exposure catalogue were corrected for geometric distortion using the publicly available WFC3/IR correction from \citet{Anderson2016}\footnote{publicly available at \url{https://www.stsci.edu/~jayander/WFC3/}}.

In addition to the positions and magnitudes, \texttt{hst1pass} can output several parameters, including the quality-fit parameter, \textit{q}. This parameter is crucial as it indicates how well the flux distribution around a source matches the $e$PSF models in a given image, which we will use it to weigh multiple measurements within each epoch (see Section~\ref{magnitudes}).

\subsection{Second-Pass Photometry}

Two of our targets, WISE\,0855$-$0714 and WD\,0806$-$661B were too faint in the available images for first-pass measurements to provide reliable photometry. For these observations, we implemented a refined ``second-pass'' photometry approach inspired by the methods described in \citet{Anderson2008}, originally developed for the ACS Survey of Globular Clusters \citep{Sarajedini2007}. This strategy is optimised for detections of very faint objects, by relying on multiple exposures to find and measure fainter sources that may have been lost in the noise of individual exposures \citep{Bellini2017,Nardiello2018}.

In a first stage, we performed an initial detection pass to identify relatively bright and isolated sources across the field using a standard peak-finding algorithm. These stars provided a preliminary astrometric and photometric reference frame and were fitted with a position-dependent PSF model. Following \citet{Anderson2008}, we then conducted a second-pass photometric analysis. The flux and position of WISE\,0855$-$0714  and WD\,0806$-$661B were derived by simultaneously fitting the PSF model across all available dithered exposures, thus optimally combining the information from multiple images. The procedure was iterated several times, updating the star list and recomputing residuals until convergence was achieved. This multi-iteration, neighbour-subtracted approach significantly improved the consistency and reliability of the photometric measurements for WISE\,0855$-$0714 and WD\,0806$-$661B, reducing the impact of blending from nearby sources, and enhancing the effective signal-to-noise of the detections compared to conventional single-pass techniques.

\section{Photometric Measurements}
\label{photometry}

\subsection{Zero-Point Photometric Calibration}
\label{phot_calibration}

The extracted photometry was then calibrated to the Vega magnitude system following the method of \citet{Bedin2005} (see also \citealt{Bellini2017,Nardiello2018,Scalco2021}).
The zero-point of \textit{HST} photometry is based on a comparison between the PSF-based instrumental magnitudes measured on the \texttt{\_flt} exposures and aperture photometry measured on the \texttt{\_drz} images. The zero-point ${\rm ZP_{PSF,X}}$ for a filter ${\rm X}$ is defined as:
\begin{equation} \label{eq_ZP}
{\rm ZP_{PSF,X}} = {\rm ZP_{AP,X}} + \langle \delta m_{\rm X} \rangle
\end{equation}
where ${\rm ZP_{AP,X}}$ is the infinite-aperture Vega-magnitude zero-point provided on the WFC3 webpage\footnote{ \url{https://www.stsci.edu/hst/instrumentation/wfc3/data-analysis/photometric-calibration/ir-photometric-calibration/}}, and $\langle \delta m_{\rm X} \rangle$ represents the median magnitude difference between $m_{\rm AP(r,\infty),X}^{\tt drz}$ ---the aperture photometry measured on the \texttt{\_drz} images within a finite aperture radius ${\rm r}$ and corrected to an infinite aperture--- and the PSF-based instrumental magnitudes $m_{\rm PSF, X}^{\tt flt}$.

To compute $m_{\rm AP(r,\infty),X}^{\tt drz}$, we performed aperture photometry on the \texttt{\_drz} images with an aperture radius of 3 pixels ($\sim$0.39\,arcsec), and corrected the measurements to an infinite aperture using the encircled energy fractions tabulated on the WFC3 webpage\footnote{\url{https://www.stsci.edu/hst/instrumentation/wfc3/data-analysis/photometric-calibration/ir-encircled-energy}}. For each filter, we cross-matched stars detected in both the \texttt{\_drz}-based aperture photometry and our PSF-based photometry, and computed the 2.5$\sigma$-clipped median of $m_{\rm AP(r,\infty),X}^{\tt drz} - m_{\rm PSF,X}^{\tt flt}$. This value was adopted as our estimate of $\langle \delta m_{\rm X} \rangle$, from which we derived ${\rm ZP_{PSF,X}}$ using Equation~\ref{eq_ZP}.
We verified that the photometric calibration is consistent across all targets and epochs to within the expected $\pm$0.02\,mag accuracy of \textit{HST} absolute photometry \citep{Bedin2005}. Given the limited number of sources in each field, we adopt a single zero-point per filter, defined as the clipped mean of the derived values in each filter. These final ${\rm ZP_{PSF,X}}$ values are listed in Table~\ref{t:ZPs}.

Finally, the calibrated magnitude $m_{\rm CAL,X}$ of a star in filter ${\rm X}$ is given by:
\begin{equation} \label{eq_mag}
m_{\rm CAL,X} = m_{\rm PSF,X}^{\tt flt} + {\rm ZP_{PSF,X}}
\end{equation}
\begin{table}
    \caption{WFC3/IR Vega-magnitude zero-points derived here for our PSF-based photometry.}
    \centering
    \begin{tabular}{l c}
        \hline\hline
        Filter & Zero-point \\
        \hline
        F105W & 25.52 \\ 
        F110W & 25.96 \\
        F125W & 25.22 \\
        F127M & 23.56 \\
        F140W & 25.27 \\
        F160W & 24.59 \\
        \hline
    \end{tabular}
    \label{t:ZPs}
\end{table}

\subsection{Combining the Calibrated Magnitudes}
\label{magnitudes}

\begin{table*}
    \caption{Y-dwarf \textit{HST} photometric measurements.}
    \centering
    \footnotesize
    \resizebox{\textwidth}{!}{
    \renewcommand{\arraystretch}{1.1}
    \begin{tabular}{lccccccccc}
        \hline\hline
        Name & T$_{\text{eff}}$ & Ref. & Parallax & F105W & F110W & F125W & F127M & F140W & F160W \\
            & [K] & & [mas] & [mag] & [mag] & [mag] & [mag] & [mag] & [mag] \\ 
        \hline
        WISE\,0336$-$0143AB & 440$^\mathrm{a}$ & (a) & 99.8 $\pm$ 2.1 & 22.27 $\pm$ 0.02  & --- & 21.91 $\pm$ 0.03 & --- & --- & 21.78 $\pm$ 0.02 \\
        WISE\,0350$-$5658 & 325 & (b) & 182.86 $\pm$ 3.07 & 22.95 $\pm$ 0.03 & --- & 22.75 $\pm$ 0.03 & 21.41 $\pm$ 0.05 & 22.33 $\pm$ 0.04 & 22.23 $\pm$ 0.05 \\
        WISE\,0359$-$5401 & 443 & (c) & 73.6 $\pm$ 2.0 & 22.55 $\pm$ 0.07 & --- & 22.07 $\pm$ 0.05 & 20.87 $\pm$ 0.09 & 21.79 $\pm$ 0.03 & 22.05 $\pm$ 0.16 \\
        WISE\,0410$+$1502 & 435 & (b) & 153.01 $\pm$ 0.70 & 20.48 $\pm$ 0.02 & --- & 19.93 $\pm$ 0.02 & 18.52 $\pm$ 0.02 & 19.65 $\pm$ 0.02 & 20.02 $\pm$ 0.02 \\
        WISE\,0535$-$7500 & 496 & (c) & 77.49 $\pm$ 9.13 & 23.13 $\pm$ 0.02 & --- & 22.85 $\pm$ 0.04 & --- & 22.40 $\pm$ 0.04 & 22.62 $\pm$ 0.04 \\
        WISE\,0647$-$6232 & 405 & (b) & 110.06 $\pm$ 2.25 & 23.53 $\pm$ 0.03 & --- & 23.38 $\pm$ 0.03 & --- & --- & 23.10 $\pm$ 0.09 \\
        WISE\,0713$-$2917 & 465 & (b) & 110.11 $\pm$ 0.61 & 21.00 $\pm$ 0.03 & --- & 20.33 $\pm$ 0.02 & 19.00 $\pm$ 0.03 & --- & 20.29 $\pm$ 0.02 \\
        WISE\,0734$-$7157 & 466 & (c) & 73.35 $\pm$ 0.71 & 21.68 $\pm$ 0.03 & --- & 20.98 $\pm$ 0.02 & 19.61 $\pm$ 0.02 & --- & 20.97 $\pm$ 0.03 \\
        WD\,0806$-$661B & 343 & (d) & 51.99 $\pm$ 0.02 & 26.00 $\pm$ 0.10 & 25.77 $\pm$ 0.05 & 25.85 $\pm$ 0.08 & 24.66 $\pm$ 0.03 & --- & 25.28 $\pm$ 0.12 \\
        WISE\,0825$+$2805 & 387 & (c) & 160.22 $\pm$ 10.25 & 23.47 $\pm$ 0.04 & --- & 23.16 $\pm$ 0.03 & --- & --- & 22.73 $\pm$ 0.04 \\
        WISE\,0855$-$0714 & 285 & (e) & 439.0 $\pm$ 2.4 & 27.42 $\pm$ 0.19 & 26.31 $\pm$ 0.07 & 26.15 $\pm$ 0.15 & 24.52 $\pm$ 0.08 & --- & 23.91 $\pm$ 0.05 \\
        WISE\,1206$+$8401 & 472 & (c) & 83.92 $\pm$ 4.96 & 21.65 $\pm$ 0.04 & --- & 21.07 $\pm$ 0.02 & 19.70 $\pm$ 0.02 & --- & 21.11 $\pm$ 0.02 \\
        WISE\,1405$+$5534 & 392 & (c) & 158.2 $\pm$ 2.6 & 21.94 $\pm$ 0.04 & 21.66 $\pm$ 0.03 & 21.72 $\pm$ 0.02 & --- & 21.25 $\pm$ 0.02 & 21.54 $\pm$ 0.02 \\
        WISE\,1541$-$2250 & 411 & (c) & 172.28 $\pm$ 1.55 & 22.39 $\pm$ 0.05 & --- & 21.92 $\pm$ 0.02 & 20.43 $\pm$ 0.03 & --- & 21.79 $\pm$ 0.06 \\
        WISE\,1639$-$6847 & 405 & (b) & 212.67 $\pm$ 0.91 & 21.75 $\pm$ 0.07 & --- & 21.51 $\pm$ 0.02 & 19.99 $\pm$ 0.02 & --- & 21.14 $\pm$ 0.03 \\
        WISE\,1738$+$2732 & 450 & (b) & 133.65 $\pm$ 0.83 & 20.72 $\pm$ 0.02 & --- & 20.22 $\pm$ 0.02 & 18.88 $\pm$ 0.02 & 19.83 $\pm$ 0.02 & 20.26 $\pm$ 0.02 \\
        WISE\,1828$+$2650 & 425 & (f) & 120.00 $\pm$ 5.94 & 23.88 $\pm$ 0.03 & --- & 23.79 $\pm$ 0.02 & --- & 23.06 $\pm$ 0.12 & 22.91 $\pm$ 0.06 \\
        WISE\,2056$+$1459 & 481 & (c) & 146.44 $\pm$ 0.43 & 20.54 $\pm$ 0.02 & --- & 19.86 $\pm$ 0.02 & 18.48 $\pm$ 0.02 & 19.47 $\pm$ 0.02 & 19.86 $\pm$ 0.02 \\
        WISE\,2209$+$2711 & 355 & (c) & 165.09 $\pm$ 12.01 & 23.83 $\pm$ 0.04 & --- & 23.51 $\pm$ 0.10 & --- & 23.14 $\pm$ 0.04 & 23.16 $\pm$ 0.10 \\
        WISE\,2220$-$3628 & 450 & (b) & 93.50 $\pm$ 1.02 & 21.60 $\pm$ 0.03 & --- & 21.01 $\pm$ 0.02 & 19.67 $\pm$ 0.02 & --- & 20.96 $\pm$ 0.03 \\
        WISE\,2354$+$0240 & 347 & (c) & 130.6 $\pm$ 3.3 & 23.37 $\pm$ 0.03 & --- & 23.28 $\pm$ 0.04 & 21.81 $\pm$ 0.04 & --- & 22.94 $\pm$ 0.07 \\
    \hline\\ [-2.5ex]
    \multicolumn{10}{l}{
    \begin{minipage}{0.98\linewidth}
        \textbf{Notes.}\\
        Dashes indicate bands with no available observations.
        Parallaxes for most sources were derived in \citetalias{Fontanive2025} using the same \textit{HST} data, with the exception of WD\,0806$-$661B which comes from the \textit{Gaia} DR3 \citep{GaiaDR3} parallax of its host star, and the other 5 targets listed in Table~\ref{t:observations} which come from \citet{Kirkpatrick2021}. The temperature for WISE\,0336$-$0143 corresponds to that inferred for the primary component.\\
        $^\mathrm{a}$ Estimated effective temperature for the primary component.\\
        References for effective temperatures: (a) \citet{Leggett2023}; (b) \citet{Leggett2021}; (c) \citet{Beiler2024}; (d) \citet{Voyer2025}; 
        (e) \citet{Luhman2024}; (f) \citet{Lew2024}.
    \end{minipage}}    \end{tabular}}
    \label{t:photometry}
\end{table*}

In many \textit{HST} workflows, individual exposures in a given dataset are tied both astrometrically and photometrically to a common reference frame using stars in common (e.g., \citealp{Anderson2008,Milone2023,Marino2024,Ziliotto2025}). Photometric registration is done by adjusting flux measurements in individual images to agree as closely as possible with those in the reference frame for common stars. This procedure optimises precision by enforcing a unified instrumental magnitude system across all exposures. However, since it relies on bright stars, that might be impacted by telescope breathing effects between exposures in different ways from our much fainter and redder Y dwarfs, this might introduce systematic errors in the measured photometry of our science targets. While we aim for high levels of precision in this study, absolute accuracy is essential for establishing reliable colour–magnitude diagrams. In contrast to \citetalias{Fontanive2021} and \citetalias{Fontanive2025}, we therefore do not average instrumental magnitudes registered to a common photometric reference frame, in order to avoid calibration steps that might result in degraded accuracy for our targets. Instead, we calibrate each image independently without enforcing relative photometric agreement between exposures, providing a more reliable accuracy in the final calibrated apparent magnitudes, obtained independently from unrelated brighter neighbours with much bluer colours than our Y dwarfs.

To determine the final magnitudes of our science targets in each epoch and filter, we used a weighted average approach that accounts for the quality of each measurement within a dataset. We used the PSF quality-of-fit parameter $q$ of the Y dwarf in each exposure to assign a weight to each individual measurement as $1/q^2$, so that higher-quality measurements contribute more significantly to the final value. The zero-point calibrations derived in Section~\ref{phot_calibration} for each filter (Table~\ref{t:ZPs}) were then added to the weighted average of the instrumental magnitudes to obtain a final apparent magnitude for each dataset.

The associated uncertainties ($\sigma$) were estimated by calculating the weighted dispersion of the individual measurements around the combined dataset magnitudes. To account for the absolute photometric calibration accuracy, we also included an additional uncertainty component of 0.02\,mag (see Section~\ref{phot_calibration}) in the final error budget. This ensures that both statistical variations and calibration limitations are properly reflected in the final uncertainties.

When multiple epochs were available in a same filter, we further computed a weighted average of the obtained magnitudes for each dataset to determine a final photometric value. The magnitude from each epoch was weighted by the inverse of its variance ($w=1/\sigma^2$), giving greater influence to more precise datasets. 
The uncertainties on the combined magnitudes were derived from the total weight of the measurements as $\sqrt{1/\sum(w)}$, ensuring that it reflects the precision of the available epochs of observation in a given filter. This approach optimally combines all available observations, while accounting for variations in uncertainties between datasets. Our final photometric measurements and their associated uncertainties in each available filter are provided in Table~\ref{t:photometry}.

Our final magnitude uncertainties are around 0.02--0.05 for the majority of targets, across all filters, with a few cases reaching errorbars around 0.1\,mag for lower S/N targets. For the two fainter sources that required a second-pass photometric analysis, WD\,0806$-$661B and WISE\,0855$-$0714, we achieved precisions of $\sim$0.05--0.1\,mag in most bands, and up to 0.15--0.2\,mag in F105W and F125W for WISE\,0855$-$0714. Measurements for these two targets are in very good agreement with literature values \citep{LuhmanEsplin2016,Schneider2016,Leggett2017}.

\section{Examinations of Y-dwarf Fluxes}
\label{results}

\subsection{Inter-epoch consistency}
\label{variability}

For targets with multiple epochs of observations available in the same filter, we explore inter-epoch consistency by computing differential magnitudes relative to the mean, alongside statistical indicators like the root-mean-square (rms) and reduced chi-squared ($\chi^2_\nu$) around the mean, to test whether the observed scatter is consistent with the measurement uncertainties. We observe very low variability across most targets, with typical values of rms $<$0.02--0.05\,mag and $\chi^2_\nu$ around $\sim$1 obtained in the majority of cases, confirming high photometric stability across epochs.

Several objects, however, exhibit more significant inter-epoch scatter. These include WISE\,1828$+$2650 in F105W and F125W, WISE\,2354$+$0240 in F125W, and WISE\,0336$-$0143 in F105W, all showing rms values in the range of $\sim$0.06--0.14\,mag and $\chi^2_\nu \sim 4$--$13$, for differences between the measured epoch magnitudes at the level of $\sim$0.1--0.3\,mag. While these variations may be indicative of genuine photometric variability (e.g., WISE\,1405$+$5534 and WISE\,1738$+$2732 are known to be variable at longer wavelengths in the mid-infrared; \citealp{Cushing2016,Leggett2016b}), we note that these sources are among the faintest in our sample and their low signal-to-noise in near-infrared \textit{HST} measurements could contribute to the increased observed scatter (see \citetalias{Fontanive2025}). In the case of the brighter WISE\,0336$-$0143, known from  \textit{JWST} to be a tight binary \citep{Calissendorff2023} on a sub-pixel separation for the WFC3/IR camera, it is possible that a change in the orbital configuration of the unresolved components over the timeline of the available F125W epochs resulted in a varying flux contribution from the faint secondary, although the latter likely lies close to the sensitivity of these \textit{HST} observations.

The most striking case is WISE\,1639$-$6847, whose two epochs of F125W data (2013.13 and 2013.83) yield rms = 0.12\,mag and $\chi^2_\nu = 44$, with an unusually high degree of variability of 0.23\,mag that stands well above that expected from measurement uncertainty alone, especially since WISE\,1639$-$6847 lies towards the brighter end of our sample. This target was precisely characterised astrometrically in our previous work \citep{BedinFontanive2020,Fontanive2021}, making the observed photometric discrepancies initially surprising. However, the precision of the 2013.83 epoch (with data acquired in F105W and F125W) was degraded compared to other datasets, due to a combination of scattered Earthlight contamination at low limb angles, shorter total exposure times leading to reduced signal‑to‑noise, and the close proximity of a relatively bright star at $\sim$3.5 pixels from the target \citep{BedinFontanive2020}. These factors strongly suggest that atmospheric variability is unlikely to be the source of this temporal discrepancy in F125W, and instead reflect limitations in the affected dataset. This degraded quality is accounted for via the quality‑of‑fit parameter $q$ (0.2--0.7 for the 2013.83 measurements, compared to $\sim$0.05 in the 2013.13 data), which naturally assigns more weight to the higher‑quality measurements from the 2013.13 epoch in the final combined F125W photometry. While the F127M and F160W data were obtained at later epochs and did not suffer from these issues, the only available dataset in F105W was obtained during the 2013.83 epoch and may be compromised by similar issues, which is reflected in the somewhat inflated uncertainty (for the target's brightness) of 0.07\,mag in the obtained F105W magnitude. We also note that \citet{Schneider2015} found F105W and F125W magnitudes 0.3--0.4\,mag brighter than ours from aperture photometry performed on the 2013.83 observations, with a background level estimated by performing aperture photometry to random star-free positions in the field of view, which may be affected by the multiple issues encountered in these datasets. When performing synthetic photometry from WFC3 grism spectra, \citet{Schneider2015} obtained a fainter F105W magnitude, within 0.1\,mag of our measurement, but a F125W flux still 0.3\,mag brighter than ours, although 0.1\,mag fainter than with aperture photometry.

\subsection{The Y-dwarf sequence in colour–magnitude diagrams}
\label{CMDs}

\begin{figure*}
    \centering
    \includegraphics[width=0.99\linewidth]{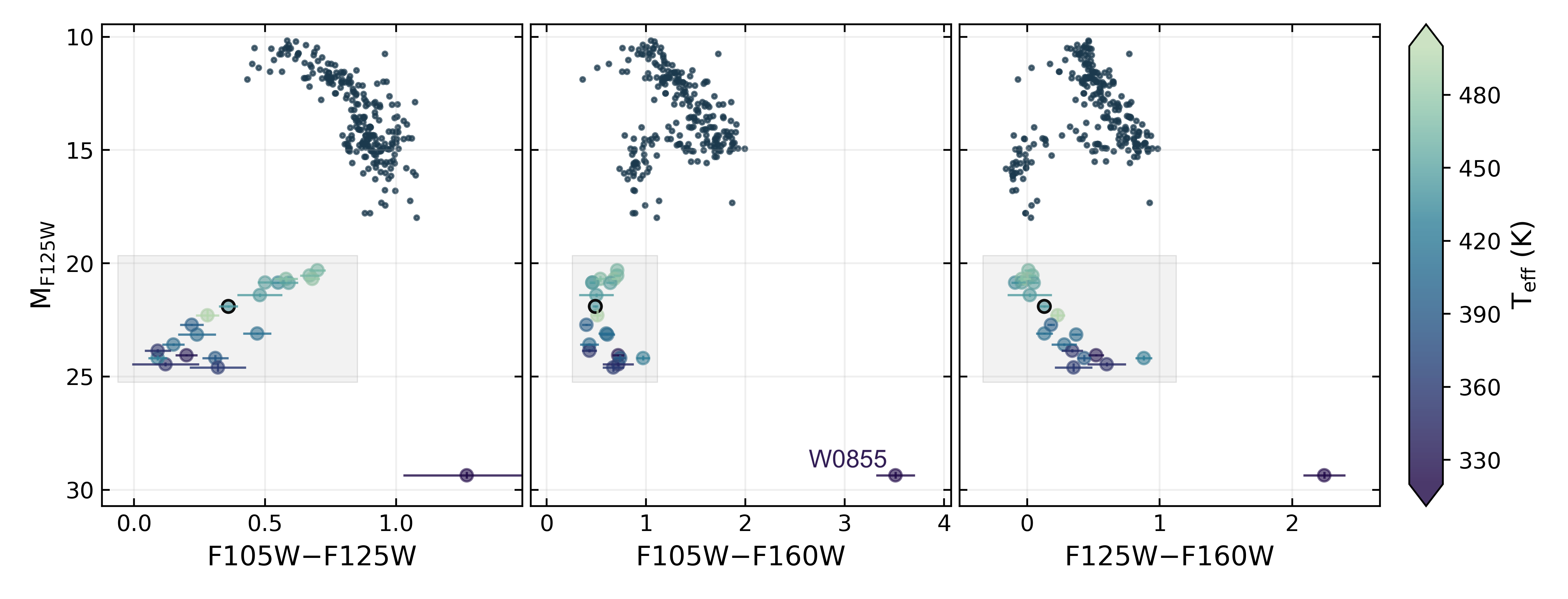}
    \includegraphics[width=0.99\linewidth]{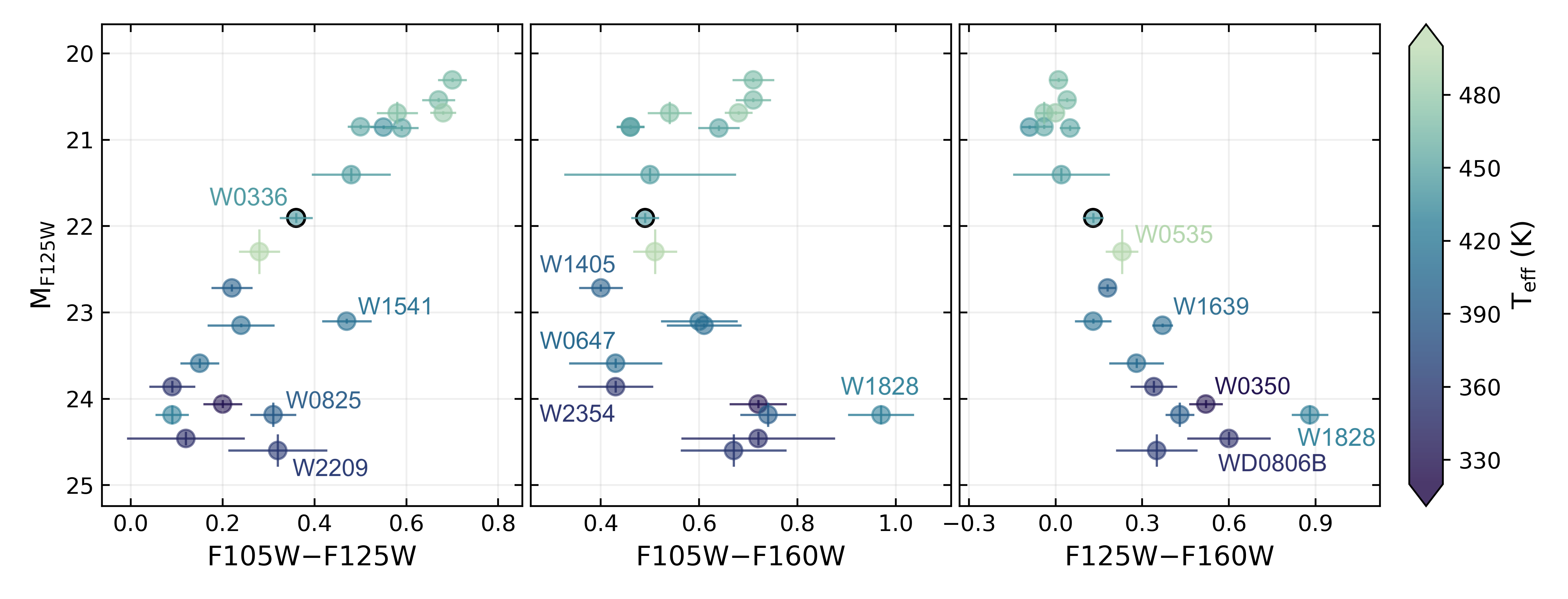}
    \caption{\textit{HST} CMDs for our 21 Y dwarfs, showing all combinations of WFC3/F105W, F125W and F160W colours against F125W absolute magnitude. The target with a black edge is the known binary WISE\,0336$-$0143 \citep{Calissendorff2023}, unresolved in \textit{HST} observations. The effective temperatures in the colour bar are reported in Table~\ref{t:photometry}. The grey points in the top panel show synthetic photometry for L and T dwarfs from the SpeX Prism Spectral Library \citep{SpeX2014}. The bottom panel shows zoomed-in views of the regions within the grey boxes in the top panels.}
    \label{f:CMD_NIR-wide}
\end{figure*}

In this section, we explore the photometric properties of the Y-dwarf population through examinations of colour–magnitude and colour–colour diagrams. We investigate various combinations of NIR CMDs from our measured multi-band \textit{HST}/WFC3 photometry in Table~\ref{t:photometry}, along with \textit{Spitzer} [3.6] and [4.5] mid-infrared photometry compiled in \citet{Kirkpatrick2021}, and \textit{JWST} data from \citet{Albert2025}. We use our recent \textit{HST}--\textit{Gaia} parallax values from \citetalias{Fontanive2025} for the 15 our targets presented in that paper. For WD\,0806$-$661B, we use the \textit{Gaia} DR3 parallax \citep{GaiaDR3} of its white dwarf host, and adopt the \textit{Spitzer}-derived parallax measurements from \citet{Kirkpatrick2021} for the remaining 5 sources (see Table~\ref{t:photometry}), in order to convert from apparent to absolute magnitudes.

\subsubsection{\textit{HST} colour–magnitude diagrams}
\label{NIR_CMDs}

\begin{figure*}
    \centering
    \includegraphics[width=0.99\linewidth]{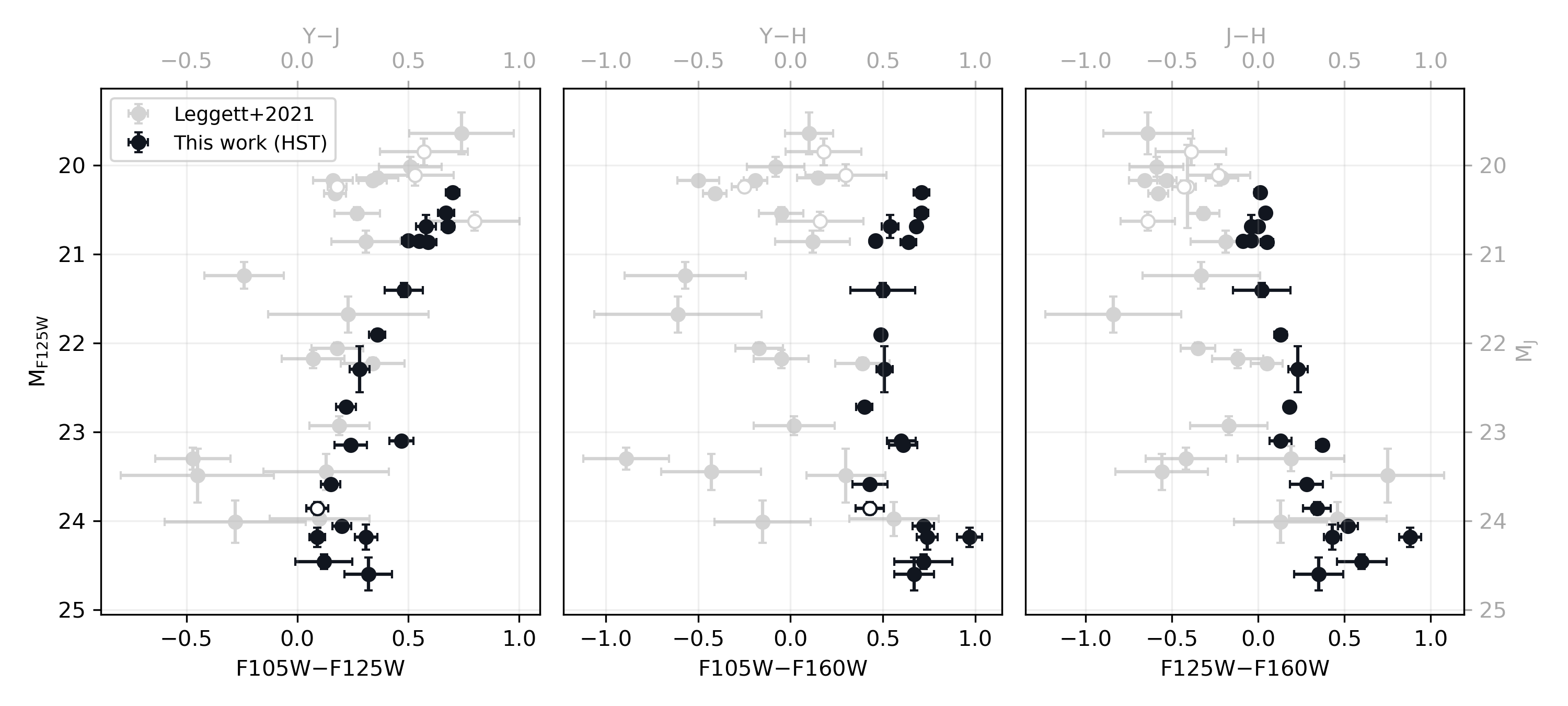}
    \caption{Comparison of the improved tightness in NIR CMDs for Y dwarfs from our \textit{HST} observations (black) relative to the compilation of \textit{YJH} MKO photometry from \citet{Leggett2021} (grey). Filled symbols mark objects present in both samples, and only a few open symbols indicate objects missing from one of the datasets: WISE\,2354$+$0240 from our sample which lacks $Y$-band photometry (left and middle panels), and 4 Y dwarfs at the bright end of the ground-based sample that are not part of our \textit{HST} sample. While plotted on the same scales for direct comparisons of the local scatters and size of uncertainties, it is important to note that axes correspond to different filter systems, and magnitude or colour values should not be directly compared between datasets.}
    \label{f:CMD_YJH}
\end{figure*}

Figures~\ref{f:CMD_NIR-wide} and \ref{f:CMD_YJH} present NIR CMD sequences from \textit{HST}, mapping all combinations of colour indices in the three WFC3 wide filters available for all targets: F105W, F125W, and F160W, against absolute magnitude in F125W. The overall trends in colour–magnitude space remain consistent across different choices of y-axis filter, and we choose to present CMDs using M$_\mathrm{F125W}$ absolute magnitude as the representative band, for consistency with prior studies, which typically show $J$-band absolute magnitudes owing to the higher completeness and uniformity of data in that band (e.g., \citealp{Tinney2014,Schneider2016,Leggett2015,Leggett2016,Leggett2017}).

In Figure~\ref{f:CMD_NIR-wide}, each Y-dwarf data point is colour-coded by the effective temperatures (T$_\mathrm{eff}$) reported in Table~\ref{t:photometry}, inferred from \textit{JWST} spectra when available, and from previous works otherwise. 
We additionally compute and incorporate synthetic photometry in the same \textit{HST} bands for all field brown dwarfs available from the SpeX Prism Spectral Library \citep{SpeX2014} and curated with parallax measurements in the \texttt{SPLAT} toolkit \citep{SPLAT2017}. Shown as grey points in the top panels, these data extend the sequence to earlier L and T spectral types, providing a wider context and baseline for the photometric evolution of Y dwarfs with temperature. The notable gap between the cool end of the SpeX spectral sequence and the warmest Y dwarfs in our sample arises because the sources observed with SpeX do not extend all the way to down the Y dwarf regime, leaving a spectral and photometric gap between the latest T dwarfs plotted here and the onset of the Y sequence. A second discontinuity is observed, between the core of our Y dwarf sequence and the position of WISE\,0855$-$0714. While recent discoveries \citep{Marocco2019,BardalezGagliuffi2020,Meisner2020} have begun to populate this region, the faintness and lack of sufficient astrometric baselines for these colder targets excluded them from selection for our \textit{HST} program, leaving this region unpopulated in our current dataset.

To further highlight the photometric behaviour of Y dwarfs with greater precision, we present in the bottom panels of Figure~\ref{f:CMD_NIR-wide} a zoomed-in view of the same CMDs, restricted to the compact region occupied by the core of our Y-dwarf sequence. The exceptionally faint and red WISE\,0855$-$0714 is excluded from this inset in order to focus on the clustered population of warmer Y dwarfs that form our core sample. Within this magnified frame, we annotate specific targets to emphasise individual benchmarks and outliers. This visualisation underscores the coherence of the CMD sequences and the subtle variations in colour and absolute magnitude that emerge across the range of temperatures represented in our sample. As seen in these diagrams, the Y-dwarf population forms a well-defined sequence in all \textit{HST} colour indices, displayed with minimal scatter, particularly in F105W--F125W and F125W--F160W. 

In F105W--F125W (Figure~\ref{f:CMD_NIR-wide}, left panel), the sequence becomes progressively bluer with fainter F125M absolute magnitude, indicating that cooler Y dwarfs emit relatively more flux in the shorter-wavelength $Y$-band range compared to the $J$-band region. This $J$--$Y$ colour trend was previously noted both in \textit{HST} studies \citep{Schneider2015,Schneider2016} and ground-based work \citep{Liu2012,LuhmanEsplin2016,Leggett2015,Leggett2021,Leggett2025}, but emerges here with markedly improved precision as seen in Figure~\ref{f:CMD_YJH}, sharpening the definition of this cooling sequence down to $\sim$350\,K. At colder temperatures, the extremely red F105W--F125W colour of WISE\,0855$-$0714 implies that a reversal in this trend must occur at some point. As suggested by \citet{Schneider2016}, we may be witnessing the onset of this reversal at the bottom of our core sequence, where the coolest sources might begin to shift back towards redder colours (WD\,0806$-$661B, WISE\,0350$-$5658, WISE\,0825$+$2805, WISE\,2209$+$2111). Pinpointing the location and shape of this colour reversal will require additional detections and observations of sources extending our sample towards WISE\,0855$-$0714 (e.g., WISE\,1446$-$2317, \citealp{Meisner2020}; WISE\,0830$+$2837, \citealp{BardalezGagliuffi2020}; WISE\,0336$-$0143B, \citealp{Calissendorff2023}; WISE\,1935$-$1546B, \citealp{DeFurio2025}). 

In contrast, the F125W--F160W colours (Figure~\ref{f:CMD_NIR-wide}, right panel) show a clear redward progression with decreasing temperature, reflecting a gradual shift in the flux distribution between the $J$- and $H$-band peaks as Y dwarfs cool. This trend had been noted in many previous studies \citep{Cushing2011,Cushing2014,Kirkpatrick2012,Schneider2015,Leggett2016,Leggett2021}, but appears here with reduced scatter compared to previously observed $J$--$H$ vs. $M_J$ CMDs for Y dwarfs (Figure~\ref{f:CMD_YJH}, see also \citealp{Schneider2016,Leggett2016,Leggett2017,Leggett2025}). This tendency does not show any indication of a reversal; instead, it likely continues gradually across the remaining gap to WISE\,0855$-$0714, whose extremely red colour indicates a monotonic continuation of this redward trend \citep{LuhmanEsplin2016,Schneider2016}. Besides WISE\,0855$-$0714, our reddest source in these F125W--F160W colours is WISE\,1828$+$2650. Interestingly, this object is also our bluest target in F105W--F125W (with WISE\,2354$+$0240), suggesting a low $J$-band flux relative to the $Y$ and $H$ bands. 

While the $Y$- and $H$-band peaks both exhibit increasingly enhanced fluxes relative to the $J$-band as temperature decreases, the F105W--F160W colour index (Figure~\ref{f:CMD_NIR-wide}, middle panel) reveals a less pronounced trend across our sequence. Even so, Figure~\ref{f:CMD_YJH} indicates that its scatter is significantly narrower than the corresponding $Y$--$H$ sequence as observed in current ground-based data. We note a tentative blueward shift in the higher temperature regime, followed by a possible inflection point near $\sim$400\,K, and a reversal towards redder colours at the cold end of our core sample. The extremely red F105W--F160W colour of WISE\,0855$-$0714 compared to the rest of the sequence indicates that the redward trend must continue at colder temperatures.

The observed NIR sequences follow the progression with decreasing effective temperature indicated in the colour bar in Figure~\ref{f:CMD_NIR-wide}, with the exception of two targets that appear significantly misplaced along the cooling sequence: WISE\,0535$-$7500 (496\,K; \citealp{Beiler2024}) and WISE\,1828$+$2650 (425\,K; \citealp{Lew2024}). Both objects have previously been flagged as suspected binaries \citep{Beichman2013,Leggett2013,Tinney2014,Leggett2017,Kirkpatrick2019,Cushing2021,Lew2024,Albert2025}, although they remain unresolved to date \citep{Opitz2016,DeFurio2023}. The effective temperatures in Table~\ref{t:photometry} consider these objects to be single, but we note that temperature estimates assuming equal-mass binaries (WISE\,0535$-$7500: 375--415\,K, \citealp{Leggett2017,Leggett2021}; WISE\,1828$+$2650: 350--380\,K, \citealp{Cushing2021,Leggett2021,Leggett2023,Barrado2023}) place them in much better agreement with the rest of the cooling sequence for their respective CMD positions. \citet{Beiler2024} noted the high temperature inferred for WISE\,0535$-$7500 from its bolometric luminosity to break the near-infrared spectral type ordering, although retrieval-based analyses of the same \textit{JWST} spectrum result in a lower effective temperature ($\sim$400\,K, \citealp{Tu2025,Lueber2026}).

One marginal outlier to the primary sequences in these CMDs is WISE\,1541$-$2250, with a F105W--F125W colour redder by $\sim$0.2--0.3\,mag compared to the rest of the sequence for its F125W absolute magnitude. Our measured F105W apparent magnitude for this target is about 0.2\,mag fainter than that from \citet{Schneider2015}, while the F125W magnitude is consistent within 0.05\,mag. \citet{Schneider2015} derived measurements from aperture photometry on \texttt{\_drz} stacks produced with \texttt{AstroDrizzle}, whereas our analysis is based on PSF fitting performed on the \texttt{\_flt} exposures. Differences between methods could account for observed offset, although apart from WISE\,1639$-$6847 (see Section~\ref{variability}), all other sources from our sample that have F105W (8 objects) or F125W (7 objects) aperture photometry measurements in \citet{Schneider2015} are in agreement with ours to $<$0.05\,mag. \citet{Schneider2015} also synthesised integrated-light photometry from WFC3 grism spectra, reporting magnitudes $\sim$0.5\,mag fainter in both bands than with aperture photometry. The source of these discrepancies remains uncertain, but we note that \citet{Leggett2013} also found WISE\,1541$-$2250 to stand out from other Y dwarfs in NIR \textit{YJH} colour trends.

To place our results in the context of previous observations of the Y-dwarf NIR sequences, we show in Figure~\ref{f:CMD_YJH} a comparison of our \textit{HST} measurements with the large set of \textit{YJH} photometry and astrometry compiled by \citet{Leggett2021}. We emphasise that although these MKO filters probe similar wavelengths to the \textit{HST} bands, their throughputs differ, and the precise locations and extents of the CMD sequences should not be compared directly. Nonetheless, the sizes of the error bars and scatter in the observed sequences can be meaningfully compared, especially given that the overlap between the two samples is nearly 1:1, with only a handful of objects (shown with open symbols) missing from one dataset or the other. As seen in these diagrams, our \textit{HST} data reveal particularly tight and well-defined NIR Y-dwarf sequences, characterised by improved measurement uncertainties and significantly smaller scatter relative to existing ground-based datasets for the same targets.

\begin{figure*}
    \centering
    \includegraphics[width=0.99\linewidth]{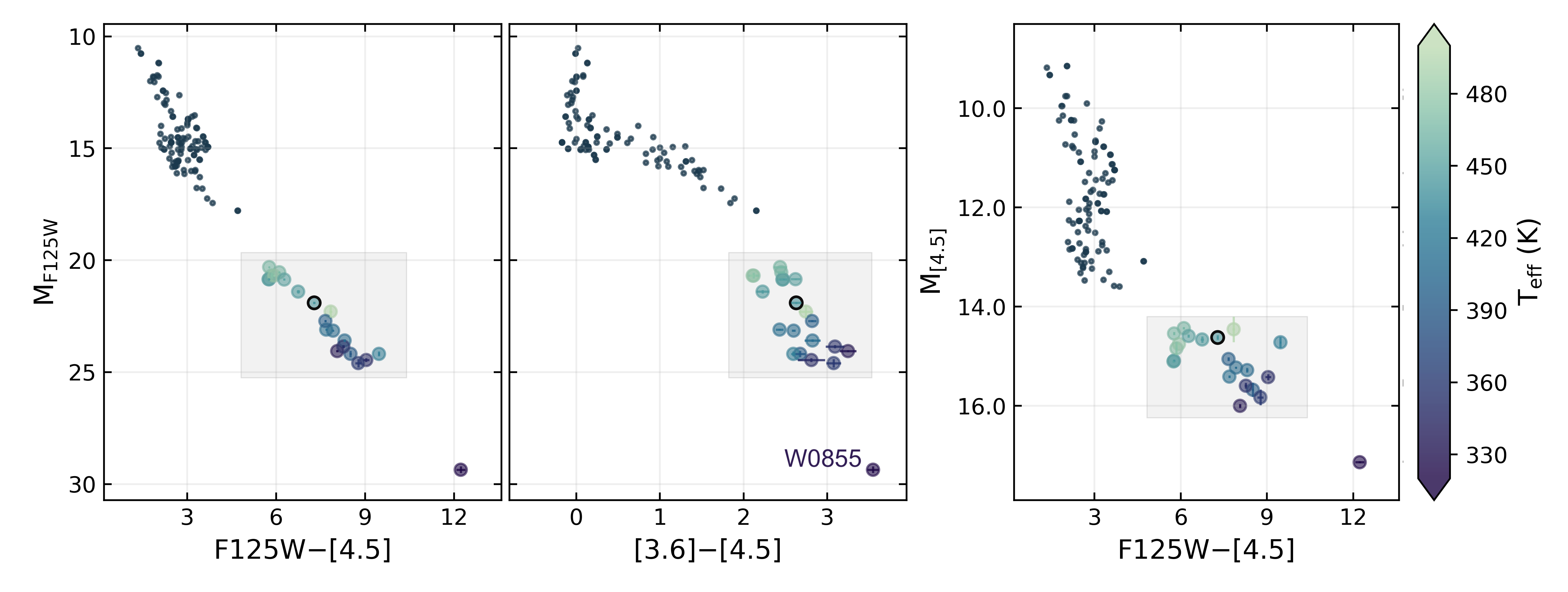}
    \includegraphics[width=0.99\linewidth]{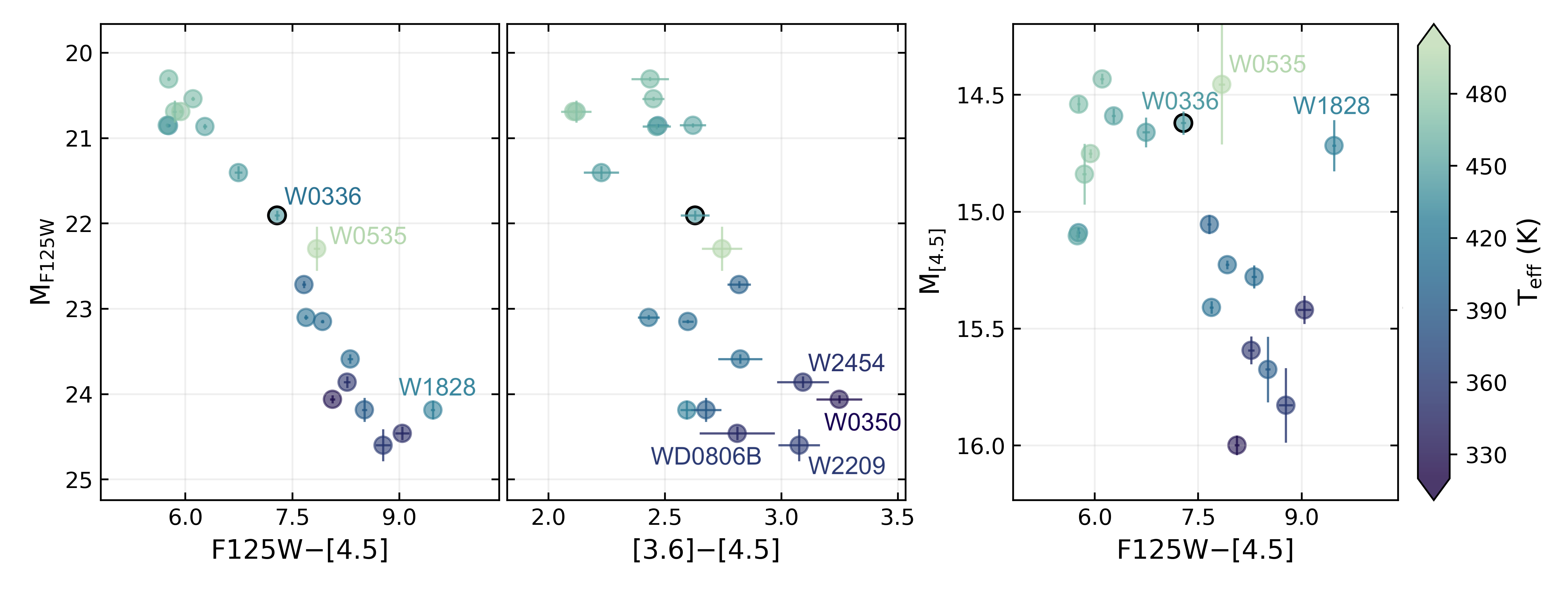}
    \caption{\textit{HST}--\textit{Spitzer} CMDs for our 21 Y dwarfs, showing WFC3/F125W absolute magnitudes against F125W--[4.5] and [3.6]--[4-5] colours (first two panels), and IRAC/[4.5] absolute magnitudes against F125W--[4.5] colours in the right-most panel. \textit{Spitzer} data come from \citet{Kirkpatrick2021}. The colour bar and symbols are the same as in Figure~\ref{f:CMD_NIR-wide}, with bottom panels showing zoom-in views of the shaded regions in the top panels.}
    \label{f:CMD_NIR-MIR}
\end{figure*}

For the two monotonic sequences ($Y$--$J$ and $J$--$H$), we tested correlations in the CMDs using the Spearman's rank correlation coefficient, which measures the strength of a monotonic relationship, adopting a Monte-Carlo approach to propagate measurement uncertainties. The obtained coefficients for our \textit{HST} data are: $-0.84\pm0.05$ for F105W--F125W, and $0.87\pm0.06$ for F125W--F160W, where values of $1$ ($-1$) indicate perfect positive (negative) correlation, and 0 means no correlation. In comparison, the data from \citet{Leggett2021} give: $-0.61\pm0.10$ for $Y$--$J$, and $0.51\pm0.11$ for $J$--$H$, confirming the tighter correlation in the \textit{HST} sequences. While the F105W--F160W sequence is not monotonic and cannot be tested in this way, Figure~\ref{f:CMD_YJH} shows that our \textit{HST} data also significantly improves the definition of the $Y$--$H$ vs. $M_J$ observational sequence.

This improvement stems from the high precision nature of our \textit{HST} photometry and astrometry, making our dataset largely free from the typical limitations imposed by uncertain distances and large photometric noise. The uniformity of our measurements, obtained on a unique instrument and extracted with the same data analysis methods, also avoids additional uncertainties introduced when heterogeneous datasets from various instruments must be colour‑corrected onto a common filter system (e.g., \citealp{Liu2012,Leggett2015,Leggett2017}), or from possible inconsistencies between photometry from imaging and integrated-light synthetic photometry from spectra (e.g., \citealp{Schneider2015,Beiler2024}). As a result, our internally consistent measurements carry very small uncertainties and show minimal scatter, revealing clear underlying CMD sequences that emerge with unprecedented tightness in all combinations of the F105W, F125W and F160W colour indices.

\begin{figure*}
    \centering
    \includegraphics[width=0.99\linewidth]{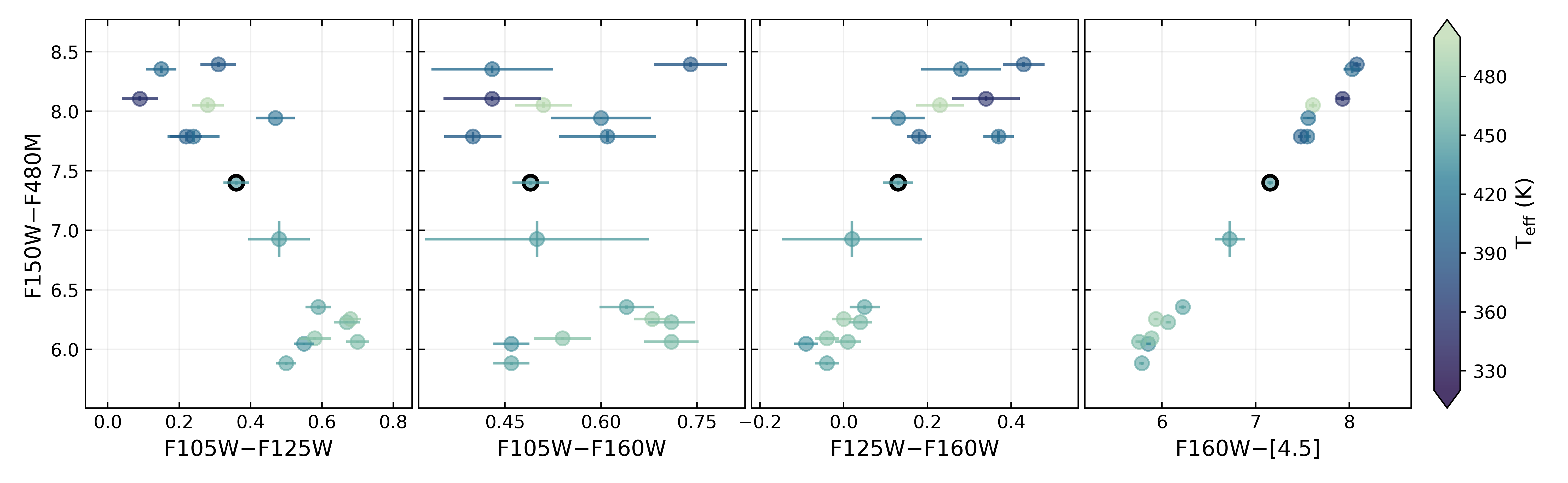}
    \caption{\textit{HST} vs. \textit{JWST} colours for the 16 of our 21 Y dwarfs with uniform NIRCam photometry from the data in \citet{Albert2025}. The first three panels display all combinations of colour indices in our main WFC3 wide bands against NIRCam F150W--F480M colours, and the rightmost panel shows the close 1:1 relationship between the \textit{JWST} colours and  F160W--[4.5] from \textit{HST} and \textit{Spitzer} covering equivalent spectral regions.}
    \label{f:CCD_HST-JWST}
\end{figure*}

\subsubsection{\textit{HST}--\textit{Spitzer} colour–magnitude diagrams}
\label{MIR_CMDs}

In Figure~\ref{f:CMD_NIR-MIR}, we examine CMDs constructed from combinations of \textit{HST} F125W data, and \textit{Spitzer} [3.6] and [4.5] photometry. As in Figure~\ref{f:CMD_NIR-wide}, the top row shows a zoomed-out view of the Y-dwarf population in these CMDs in the context of warmer brown dwarfs, and the bottom row displays insets of the shaded boxes. Field sources selected from \texttt{SPLAT} \citep{SPLAT2017} for synthetic \textit{HST} photometry were crossmatched with the Ultracool Sheet catalogue \citep{Best2020} to retrieve available \textit{Spitzer} data for the same objects; however, only a limited subset of sources had corresponding \textit{Spitzer} measurements in the catalogue (grey scatter points).

By mimicking the broad shape of the spectral energy distribution (SED) from 1 to 5\,$\mu$m, the $J$--[4.5] colour (along with other NIR--MIR indices; e.g., \citealp{Kirkpatrick2019}) is a good proxy for effective temperature for ultracool T and Y dwarfs, and forms a well-defined sequence against $M_J$ \citep{Faherty2014,Leggett2013,Leggett2016,Leggett2021,BardalezGagliuffi2020,Martin2018,Meisner2020}, although with some dependence on gravity, clouds and metallicity \citep{Tinney2014,Schneider2015,Leggett2017}. The left panel of Figure~\ref{f:CMD_NIR-MIR} illustrates this strong monotonic progression towards redder F125W--[4.5] colours with fainter $M_\mathrm{F125W}$ (see also \citealp{Schneider2016}), which closely follows the temperature sequence indicated by the colour bar (for typical uncertainties of $\pm$20--30\,K) with the exception of the two suspected binaries discussed in Section~\ref{NIR_CMDs}.

While the F125W--[4.5] colours map tightly onto $M_\mathrm{F125W}$, the second panel of Figure~\ref{f:CMD_NIR-MIR} shows that the trend towards redder [3.6]--[4.5] colours for colder objects \citep{Kirkpatrick2012,Kirkpatrick2019,Meisner2020} exhibits a less tightly-defined sequence, with a large range of absolute F125W magnitudes for the same colours, and a larger local scatter than seen in the NIR sequences (Figure~\ref{f:CMD_NIR-wide}). The enhanced intrinsic scatter in [3.6]--[4.5] (see also \citealp{Kirkpatrick2012,Martin2018}) is likely due to the strong sensitivity of the methane absorption feature at 3.6\,$\mu$m to variations in metallicity, surface gravity, and disequilibrium chemistry (e.g., \citealp{Leggett2013,LacyBurrows2023,Meisner2023}), causing the [3.6] flux to vary more strongly than other bands under the influence of secondary parameters. As in Section~\ref{NIR_CMDs}, we quantify the strength of the correlation in these monotonic CMD sequences with the Spearman's rank, which yields: $0.93\pm0.01$ for F125W--[4.5] vs. $M_\mathrm{F125W}$, and $0.68\pm0.05$ for [3.6]--[4.5] vs. $M_\mathrm{F125W}$. This confirms that F125W--[4.5], similarly to the NIR F105W--F125W and F125W--F160W colours, tracks absolute F125W magnitude closely as temperature decreases, whereas the reddening of the [3.6]--[4.5] sequence shows a weaker correlation.

Apart from WISE\,0855$-$0714, WISE\,1828$+$2650 (425\,K; \citealp{Lew2024}) is our reddest source in F125W--[4.5] colours, and stands out with more extreme colours than other sources of similar or fainter absolute F125W magnitudes, that all have lower estimated effective temperatures.
Together with WISE\,0535$-$7500 (496\,K; \citealp{Beiler2024}), these two sources appear slightly overluminous in the F125W--[4.5] CMD plotted against F125W absolute magnitude (Figure~\ref{f:CMD_NIR-MIR}, left panel), though they fall in line with other Y dwarfs when looking at [3.6]--[4.5] colours (middle panel). This elevated luminosity becomes much more pronounced when plotted against [4.5] absolute magnitude (right panel), where both objects occupy an overly bright regime for their observed NIR--MIR colours, more in line with their effective temperatures inferred from spectra covering the MIR region. Interestingly, the one confirmed binary in our sample, WISE\,0336$-$0143\,AB \citep{Calissendorff2023}, while less obviously offset in the left panel, falls in the same overluminous region in the $M_{[4.5]}$ diagram, reinforcing the suspicion that WISE\,0535$-$7500 and WISE\,1828$+$2650 are likely unresolved binaries \citep{Beichman2013,Leggett2013,Tinney2014,Leggett2017,Cushing2021,DeFurio2023}.

\subsubsection{\textit{HST}--\textit{JWST} colour–colour diagrams}

16 of our 21 targets were observed with \textit{JWST}/NIRCam as part of program GO-2473 (PI Albert). In Figure~\ref{f:CCD_HST-JWST}, we use the uniform sets of measured NIRCam photometry reported in \citet{Albert2025} for this subset of our sample to explore the relationship between the \textit{JWST}/NIRCam F150W--F480M colour and our three primary \textit{HST}/WFC3 near-infrared colour indices: F105W--F125W, F105W--F160W, and F125W--F160W (first three panels), and F160W--[4.5] (rightmost panel) which best match the wavelengths of the NIRCam filters. Each panel traces the photometric evolution of Y dwarfs across effective temperatures ranging from $\sim$475\,K down to $\sim$350\,K. 

The \textit{JWST} F150W--F480M colour, like other NIR--MIR indices, becomes progressively redder with decreasing temperature. As a result, we recover the same underlying photometric tendencies observed in the \textit{HST} NIR CMDs as a function of absolute magnitude and temperature, including the blueward progression in F105W--F125W, the redward shift in F125W--F160W, and a more complex behaviour in F105W--F160W. \citet{Albert2025} noted that the F480M absolute magnitude is a useful indicator of T$_\mathrm{eff}$, by probing the strength of the CO absorption band at $\sim$4.7\,$\mu$m with minimal sensitivity to metallicity or gravity. The F150W--F480M colours, however, show stronger on dependence on these parameters \citep{Albert2025}, reflecting the higher sensitivity of the F150W flux to secondary physical and atmospheric effects. This is consistent with the idea that the $H$--[4.5] colour is sensitive to vertical mixing \citep{Phillips2020}, and likely contributes to the enhanced scatter observed in the first three panels of Figure~\ref{f:CCD_HST-JWST} compared to our tight \textit{HST} sequences. The close wavelength correspondence between the \textit{HST} F160W and \textit{JWST} F150W bands, and between the \textit{Spitzer} [4.5] and \textit{JWST} F480M filters, produces a tight 1:1 colour–colour sequence in the rightmost panel.

\begin{figure*}
    \centering
    \includegraphics[width=0.99\linewidth]{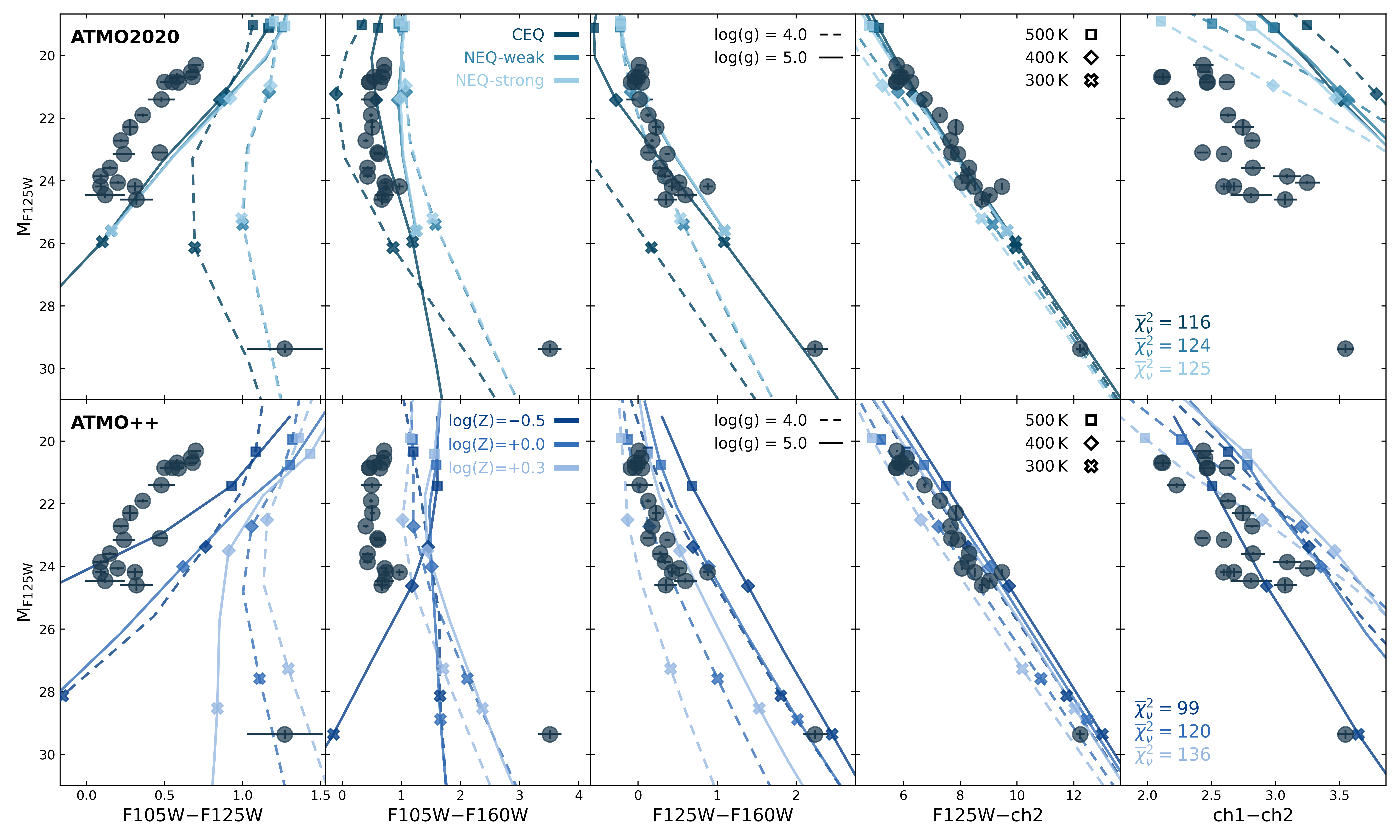}
    \caption{Comparison of the measured photometry for our 21 Y dwarfs (grey circles) to atmospheric models from the \texttt{ATMO} family \citep{Phillips2020,Leggett2021}. The different line colours correspond to grids of varying model parameters: three states of chemical (dis)equilibrium for \texttt{ATMO\,2020} (top), and metallicities of $-$[0.5], [0.0] and $+$[0.3]\,dex for \texttt{ATMO++} (bottom), with dashed and solid lines corresponding to surface gravities $\log(g)$ of 4.0 and 5.0, respectively. Squares, diamonds and crosses indicate the points along each model curve where the effective temperature T$_\mathrm{eff}$ is 500\,K, 400\,K and 300\,K, respectively. In the right-most panels, the $\overline{\chi}^2_\nu$ values represent the global reduced chi-squares of the sub-models in the corresponding colours to the full population (see Section~\ref{population-fits}.)}
    \label{f:CMD_models_ATMO}
\end{figure*}

\section{Comparisons to atmospheric models}
\label{models}

In order to investigate the physics and chemistry governing the spectro-photometric properties of Y dwarfs, we compare our measured photometry to theoretical models of substellar atmospheres. We consider three main families of atmospheric models tailored to ultracool Y-dwarf temperatures: the \texttt{ATMO}, \texttt{Lacy\,\&\,Burrows}, and \texttt{Sonora Elf\,Owl} models. Below, we provide a brief overview of each suite of model and the flavours of atmospheric parameters found within each set of model grids, and compare them to our observed CMDs. 

In the following sections, all synthetic photometry from model spectra is calculated assuming a distance of 10\,pc and a radius of 1\,R$_\mathrm{Jup}$. Colours are unaffected by these assumptions, and absolute magnitudes for our targets are adjusted to a 10-pc reference using measured parallaxes from our or other previous work. While evolved field brown dwarfs are expected to have radii close to 1\,R$_\mathrm{Jup}$, the actual radii of our targets are unknown and likely vary across the sample. Therefore, direct comparisons between observed and model absolute magnitudes should in principal also account for differences in radius. That said, any offsets introduced by radius variations are expected to be small relative to the wide dynamic range of absolute magnitudes considered, and we consider this to be negligible for visual comparisons in the CMDs below. This flux scaling factor from the radius is removed when looking at colours only, and colour–colour diagrams are unaffected by this effect.

\subsection{The ATMO models}

The \texttt{ATMO\,2020} models, developed by \citet{Phillips2020}, simulate atmospheres of self-luminous cool brown dwarfs and giant exoplanets. These models assume 1D radiative-convective equilibrium at solar metallicity, and include updated atomic and molecular opacities. The \texttt{ATMO\,2020} grids are offered with equilibrium chemistry (CEQ) and non-equilibrium chemistry with moderate (NEQ-weak) and more vigorous (NEQ-strong) vertical mixing rates.
Among all atmosphere models explored here, the \texttt{ATMO\,2020} grids are the only ones that are coupled with a self-consistent evolutionary model component. For consistency across comparisons, however, we restrict our analysis to the spectral grids alone, without incorporating the evolutionary tracks.

We also explore the \texttt{ATMO++} models \citep{Leggett2021}, an extension of the \texttt{ATMO\,2020} chemical nonequilibrium models with strong mixing. These updated models incorporate adiabat-adjusted pressure-temperature profiles, designed specifically for objects cooler than 600\,K, where standard adiabatic assumptions break down due to disrupted convection and rapid rotation. Grids of spectra for these cloud-free models are provided for solar and non-solar metallicities ($-$[0.5], [0.0] and $+$[0.3]\,dex).

In Figure~\ref{f:CMD_models_ATMO}, we compare synthetic colours and absolute magnitudes predicted by various configurations of the \texttt{ATMO\,2020} (top) and \texttt{ATMO++} (bottom) grids to the observed \textit{HST} and \textit{Spitzer} photometry of our Y-dwarf sample. All main \texttt{ATMO} sub-models successfully reproduce the observed F125W--[4.5] colours, and to some extent can account for the F125W--F160W sequence. Differences emerge, however, when considering other colour indices: the \texttt{ATMO\,2020} grids, especially those in chemical equilibrium, provide a closer match to the observed F105W--F160W colours, whereas the \texttt{ATMO++} models yield a better agreement in the MIR [3.6]--[4.5]. None of the plotted grids reproduce the observed F105W–F125W colours, which remain systematically bluer than model predictions, consistent with earlier findings that the \texttt{ATMO} models are too faint in the $Y$ band (equivalent to the \textit{HST} F105W filter) at these temperatures \citep{Phillips2020,Leggett2021}.

\subsection{The Lacy \& Burrows models}

\begin{figure*}
    \centering
    \includegraphics[width=0.99\linewidth]{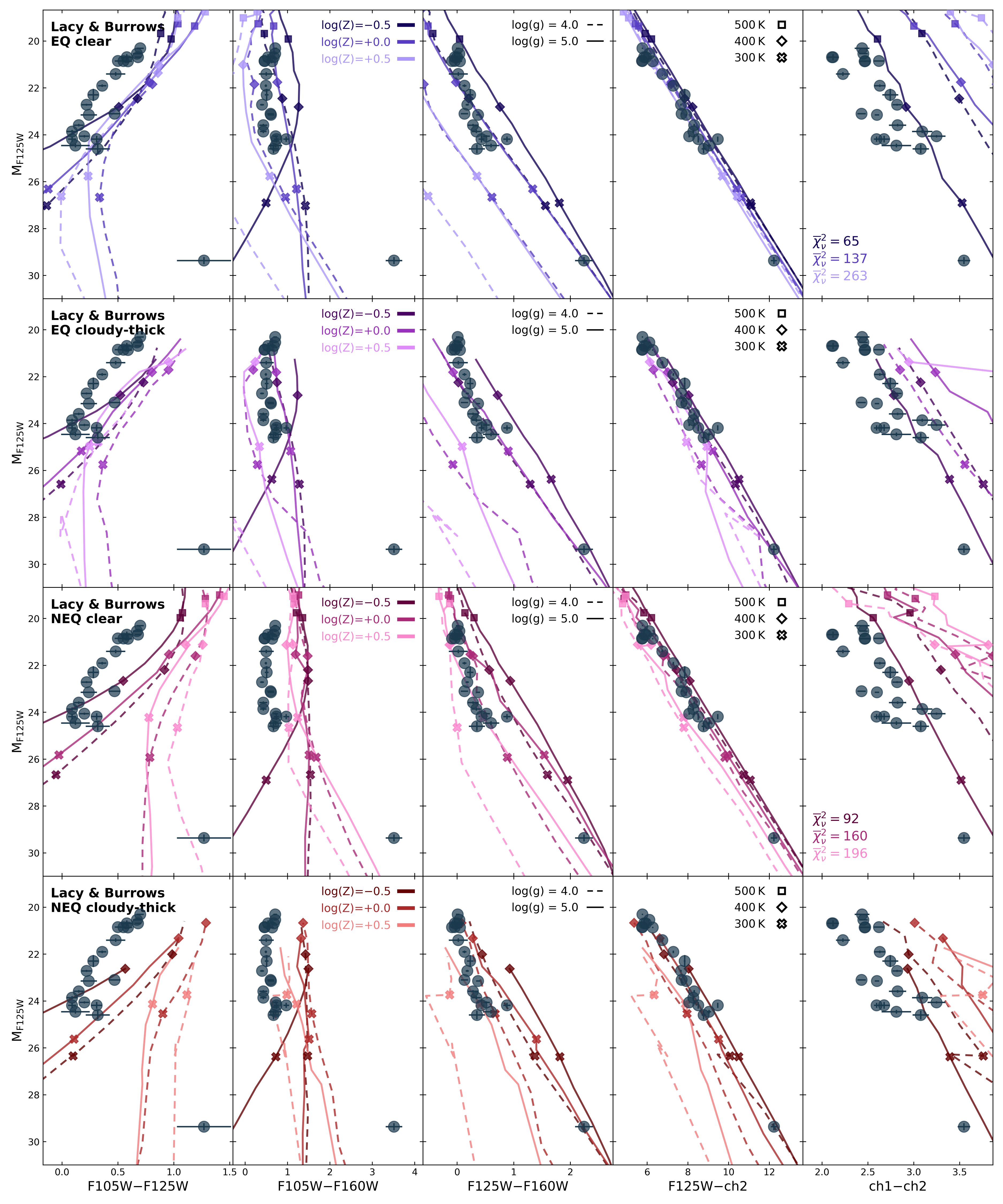}
    \caption{Same as Figure~\ref{f:CMD_models_ATMO} for the \texttt{Lacy\,\&\,Burrows} models \citep{LacyBurrows2023}. Each row shows a different combination of cloudy vs. clear atmosphere and equilibrium vs. nonequilibrium chemistry. Within each row, grids for metallicities of $-$[0.5], [0.0] and $+$[0.5]\,dex are shown from darker to lighter shades. The global fits to the population, characterised by the $\overline{\chi}^2_\nu$ values in the right-most panels, were only computed for grids in chemical equilibrium, as the grids out of equilibrium only start at T$_\mathrm{eff}$ of 400--450\,K.}
    \label{f:CMD_models_LACY}
\end{figure*}

The \texttt{Lacy\,\&\,Burrows} models consist of a comprehensive suite of self-consistent 1D radiative-convective equilibrium models tailored for Y dwarfs, developed by \citet{LacyBurrows2023}. They incorporate water clouds and disequilibrium CH$_4$--CO / NH$_3$--N$_2$ chemistry. Available grids cover a range of clear and cloudy atmospheres, both in chemical equilibrium and disequilibrium, and span metallicities of $-$[0.5], [0.0] and $+$[0.5]\,dex.

Figure~\ref{f:CMD_models_LACY} compares the observed Y-dwarf sequence against the \texttt{Lacy\,\&\,Burrows} grids of model spectra in the clear atmosphere and most optically thick cloud configurations. The tight F125W--[4.5] sequence is well reproduced by the models, with only limited variations across different physical assumptions and grid parameters. As for the \texttt{ATMO} models, most grids predict F105W--F125W and [3.6]--[4.5] colours that are systematically redder than the observed sequence, with low-metallicity curves providing the closest matches to observations. Again, models in chemical equilibrium provide a better fit to the F105W--F160W colours, whereas the non-equilibrium grids reproduce the F125W--F160W colours in a way that is less sensitive to variations in surface gravity or metallicity. The inclusion of clouds mainly affects the spread of predicted colours across metallicity and gravity, but does not appear to yield a systematically better or worse match to the overall observed sequence. 

\subsection{The Sonora Elf Owl models}

Part of the wider \texttt{Sonora} model family (\texttt{Bobcat}, \citealp{Marley2021}; \texttt{Cholla}, \citealp{Karalidi2021}), the \texttt{Elf\,Owl} Y-dwarf grids \citep{Mukherjee2023,Mukherjee2024} were designed for cloud-free atmospheres with vertical mixing-induced disequilibrium chemistry. These self-consistent 1D radiative-convective equilibrium models encompass a wide range of sub-solar to super-solar metallicities and C/O ratios, with variations in vertical mixing (characterised by the eddy diffusion coefficient $K_{zz}$) across multiple orders of magnitude. 

\begin{figure*}
    \centering
    \includegraphics[width=0.99\linewidth]{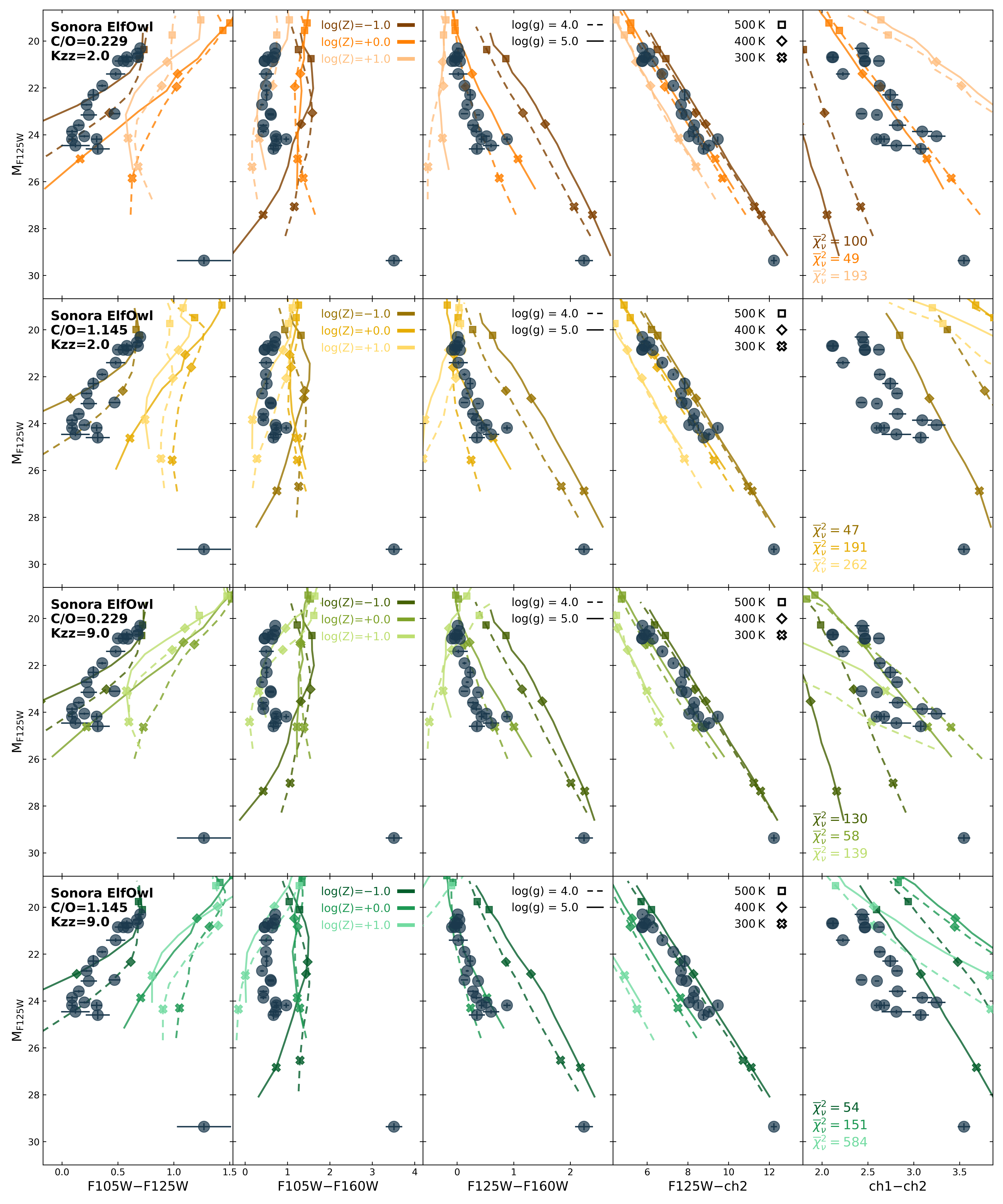}
    \caption{Same as Figure~\ref{f:CMD_models_ATMO} for the \texttt{Sonora Elf\,Owl} models \citep{Mukherjee2024}. Each row shows a different combination of the extreme values for C/O ratio (0.229 or 1.145) and vertical mixing coefficient (2 or 9\,dex). Within each row, grids for metallicities of $-$[1.0], [0.0] and $+$[1.0]\,dex are shown from darker to lighter shades. The faint and red WISE\,0855$-$0714 was excluded from the $\overline{\chi}^2_\nu$ calculations (given in the right-most panels) as these models do not extend to low enough temperatures.}
    \label{f:CMD_models_ELFOWL}
\end{figure*}

Figure~\ref{f:CMD_models_ELFOWL} shows all combinations (one per row) of the boundary values for C/O ratio (0.229 to 1.145) and vertical mixing ($\log(K_{zz})$ of 2.0 and 9.0), and within each row includes curves for sub-solar ($-$[1.0]\,dex), solar ([0.0]\,dex), and super-solar ($+$[1.0]\,dex) metallicities. The more extreme $-$[1.0]\,dex metallicity grids (darkest shades) are the only ones that reach the observed blue F105W--F125W colours of our Y-dwarf population, although higher metallicity grids provide a better match to the F105W--F160W and F125W--F160W NIR indices. In particular, across all explored families of models, the super-solar metallicity \texttt{Elf\,Owl} curves (lightest shades) are the only grids with chemical disequilibrium that can explain the observed F105W--F160W colours. The \texttt{Elf\,Owl} models also predict a pronounced dependence on metallicity in the F125W--[4.5] colours at fixed temperature, with similar slopes but a blueward shift with higher metallicity, and marginal dependence on surface gravity. This contrasts with the tightly aligned observed sequence, which shows only limited spread in colour around this slope with temperature. The effects of varying C/O ratio and vertical mixing are most apparent at MIR wavelengths, where lower C/O ratios appear to be necessary to reproduce the observed [3.6]--[4.5] locus. 

\subsection{Population-level fits to model grids}
\label{population-fits}

Since it is clear that no single model successfully reproduces all our observations, we do not attempt to constrain the properties of individual sources through SED fits to our photometry, which is better suited to dedicated spectroscopic characterisation work (e.g., \citealp{Beiler2024}). Instead, we adopt a population-level approach, focusing on relative comparisons within model families to identify which combinations of physical assumptions most closely reproduce the global photometric trends of our Y-dwarf sample. This strategy enables us to probe the dominant atmospheric processes shaping the observed characteristics of Y dwarfs, without over-interpreting fits to individual objects.

To this end, we perform a comparative chi-square analysis across multiple model grids. Using the \texttt{Dynesty} nested sampling algorithm implemented in the \texttt{species} python package \citep{Stolker2020}, we fit various grids of model spectra to the observed photometry of each target, varying effective temperature, surface gravity and a flux scaling factor (that jointly accounts for distance and radius effects), while holding all other model parameters fixed (e.g., metallicity, chemical equilibrium state, cloud coverage, C/O ratio, vertical mixing). Within each sub-model (represented by different line shades in Figures~\ref{f:CMD_models_ATMO}--\ref{f:CMD_models_ELFOWL}), we then compute a chi-square between observed absolute magnitudes and synthetic predictions from the best-fit solution to each target ($\chi^2_i$). Finally, we estimate a global reduced chi-square for the full sample, treating the whole population as a single dataset: $\overline{\chi}^2_\nu = \sum_i \chi^2_i / \sum_i \nu_i$, where $\nu_i$ is the number of degrees of freedom in each fit. By examining the aggregate chi-square obtained for various model parameters, this population-level fitting approach allows us to evaluate the goodness-of-fit of distinct model settings across the entire population, without tailoring the models to individual objects (apart from T$_\mathrm{eff}$ and $\log(g)$). 

We note that in the \texttt{Lacy\,\&\,Burrows} models, the cloudy models (both thin and thick) begin at T$_\mathrm{eff}$$<$400--450\,K and do not span the full temperature range of our warmer targets. As a result, this exercise is limited to the clear atmospheres only, and the grids that include water clouds unfortunately cannot be investigated in the same way at this stage. For the \texttt{Elf\,Owl} models, the grids do not extend to out to WISE\,0855$-$0714, and we choose to exclude it from this analysis and fit the models to the remaining 20 Y dwarfs in our sample. Given the distinct parameter space coverage of various (sub)models, the obtained global chi-square values across model families should not be interpreted as directly comparable. Our analysis is therefore restricted to internal consistency within each sub-model framework, in order to test the effects of specific parameters.

\begin{figure*}
    \centering
    \includegraphics[width=0.99\textwidth]{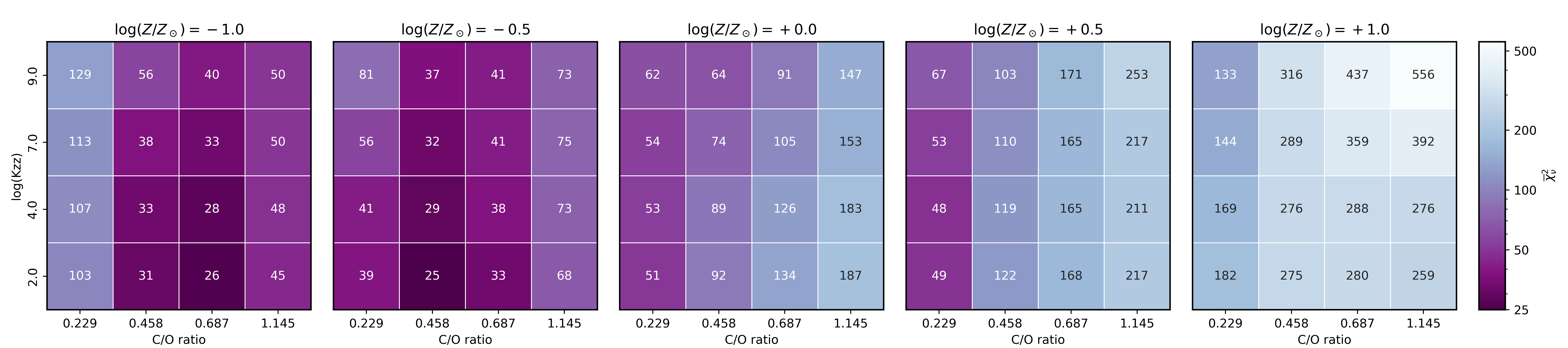}
    \caption{Global reduced chi-squared values ($\overline{\chi}^2_\nu$) for fits of the \texttt{Sonora Elf\,Owl} atmosphere models to our Y-dwarf photometry, shown as heatmaps in the C/O ratio vs. $K_{zz}$ plane. Each panel corresponds to a different metallicity, and grid cells are coloured and annotated according to their $\overline{\chi}^2_\nu$ values, with darker purple shades denoting better fits.}
    \label{f:chi2-map}
\end{figure*}

For each target, we include all available \textit{HST} flux measurements (Table~\ref{t:photometry}) along with \textit{Spitzer} [3.6] and [4.5] photometry (\citealp{Kirkpatrick2021} and references therein), in both the nested sampling fits and chi-square calculations. This results in varying degrees of freedom across the sample, but is taken into account in the weighted mean, ensuring that targets with wider photometric coverage contribute more strongly to the global chi-square. 
Final values for each sub-model are marked in the right-most panel of Figures~\ref{f:CMD_models_ATMO}--\ref{f:CMD_models_ELFOWL} for each tested sub-model. As expected from the fact that none of the models capture the reality of the observations, the resulting $\overline{\chi}^2_\nu$ are rather far from 1, varying from $\sim$50 for the best cases up to over 500 (although again these should not be directly compared across models). The main result emerging from this analysis is the indication that low-metallicity grids (darker shades) are consistently favoured across all model families, with the exception of low C/O ratio cases in the \texttt{Elf\,Owl} models, where solar-metallicity grids provide the best fits to the overall population. Interestingly, in both the \texttt{ATMO\,2020} and \texttt{Lacy\,\&\,Burrows} results, sub-models in chemical equilibrium result in better overall fits than non-equilibrium grids (for the same other model assumptions), likely due to the better match they provide to the F105W--F160W colours. Within the \texttt{ATMO} models, however, the sub-solar metallicity adiabat-adjusted \texttt{ATMO++} grids provide the lowest $\overline{\chi}^2_\nu$.

All available combinations of model settings within the \texttt{ATMO} and \texttt{Lacy\,\&\,Burrows} grids are shown in Figures~\ref{f:CMD_models_ATMO} and \ref{f:CMD_models_LACY} (apart from the thin cloud \texttt{Lacy\,\&\,Burrows} grids, which are omitted from this analysis because they do not extend to sufficiently high temperatures to encompass our warmer targets). 
In contrast, the \texttt{Elf\,Owl} model spectra span a wider range of C/O ratios, vertical mixing coefficients, and metallicities than shown in Figure~\ref{f:CMD_models_ELFOWL}. Our initial analysis focused on the boundary values of these grids, exploring the minimum and maximum settings to establish the outer limits of model behaviour. Based on the results obtained for these model configurations, we extend the same tests to intermediate grid values to complement the parameter extremes and better identify the combinations of physical parameters that most effectively reproduce the observed population. 

\begin{figure*}
    \centering
    \includegraphics[width=0.99\linewidth]{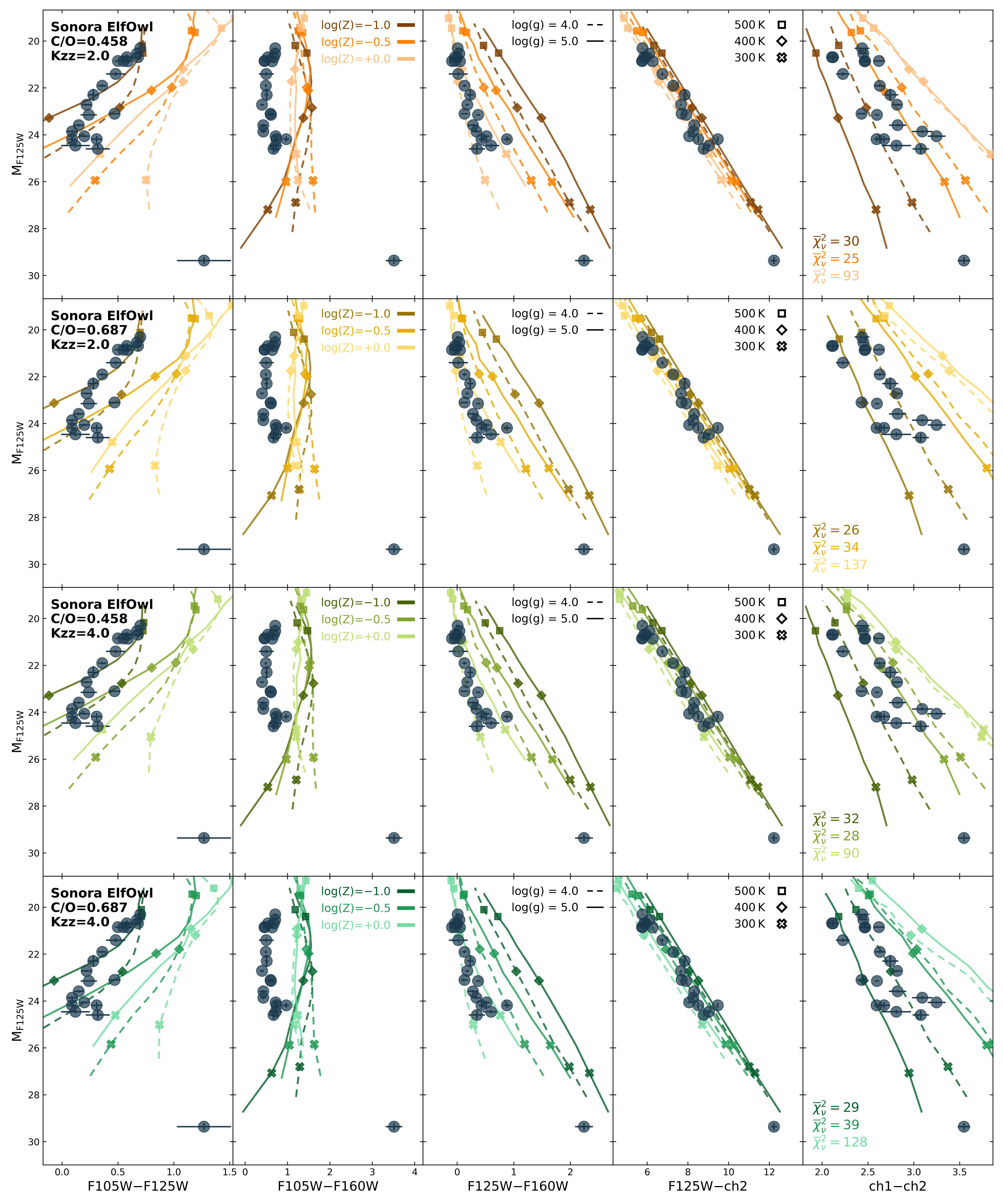}
    \caption{Same as Figure~\ref{f:CMD_models_ELFOWL}, plotting model predictions from the \texttt{Sonora Elf\,Owl} grids at intermediate C/O ratios, for the sub-solar metallicities and low vertical mixing values that provide the best overall fits to our full population (excluding WISE\,0855$-$0714).}
    \label{f:CMD_models_ELFOWL-2}
\end{figure*}

In Figure~\ref{f:chi2-map}, we show the resulting $\overline{\chi}^2_\nu$ computed in the same way for additional combinations of metallicity, C/O ratio and $K_{zz}$ available in the \texttt{Elf\,Owl} grids. These confirm that sub-solar metallicities (first two panels) are favoured, with best fits at intermediate C/O ratios (0.458--0.687 --- untested in our earlier analyses) and a preference towards lower $K_{zz}$ values. Grids from these best-fit results are shown in Figure~\ref{f:CMD_models_ELFOWL-2}.
To statistically rule out possible subsets of the parameter space, we rely on the Bayesian information criterion (BIC), for which the full (non reduced) chi-squares $\overline{\chi}^2$ are needed. Since all model fits for the \texttt{Elf\,Owl} grids have the same free parameters (T$_\mathrm{eff}$, $\log(g)$ and a flux scaling) and utilise the same sets of photometric measurements, all models have the same complexity and the $\Delta$BIC is equivalent to $\Delta\overline{\chi}^2$. With a total of 124 datapoints fitted (excluding WISE\,0855$-$0714), and 3 free parameters for each of the 20 targets considered, the reduced chi-squares in Figure~\ref{f:chi2-map} need to be multiplied by $124 - (3\times20)=64$ degrees of freedom to obtain the corresponding $\overline{\chi}^2$. This leads to a $\Delta\mathrm{BIC}=60$ between our best two values ($\overline{\chi}^2_\nu$ of 25.1 and 26.1), significantly favouring the best fit point in the \texttt{Elf\,Owl} grids over any others. With $\Delta\mathrm{BIC}\sim1500$ to the best fits at metallicities of [0.0] and $+$[0.5]\,dex ($\overline{\chi}^2_\nu\sim50$), BIC model selection overwhelmingly rules out grids with solar and super-solar metallicities, although this may not reflect a true population-level low metallicity tendency as discussed in Section~\ref{discussion} below.

\section{Discussion}
\label{discussion}

\subsection{The importance of homogenous photometry}

Our \textit{HST} data provide some of the highest precision colour–magnitude diagrams for the Y-dwarf population in the near-infrared spectral region. With uncertainties at the 0.02--0.05\,mag level for most targets, our compilation of \textit{HST} fluxes reveal a particularly well‑defined Y-dwarf locus in colour–magnitude space, with tight trends across each NIR colour index between the F105W, F125W and F160W \textit{HST} filters as effective temperature decreases. With a sample exceeding 20 targets, these sequences substantially extend previous explorations of Y‑dwarf CMDs with \textit{HST} \citep{Schneider2016} and display noticeably tighter trends than diagrams constructed from currently available ground‑based datasets \citep{LuhmanEsplin2016,Leggett2017,Leggett2021}. This improvement is driven largely by the availability of robust parallaxes, the superior signal‑to‑noise delivered by \textit{HST} at these faint flux levels, and the excellent stability achieved from space, which together minimise both vertical and horizontal uncertainties in the CMDs.

Although substantial NIR ground‑based photometry exists for Y dwarfs, these measurements are drawn from a variety of facilities whose filters differ in throughput, even for nominally similar bandpasses \citep{Liu2012}. Such differences introduce non‑negligible photometric offsets, particularly for late‑T and Y dwarfs, that can lead to large disparities between independent measurements \citep{Leggett2015}, in addition to the already larger photometric uncertainties achieved from the ground for such red and faint objects. Colour transformations can be derived and applied, but are typically based on small samples and exhibit significant intrinsic scatter for the coldest objects \citep{Liu2012,Leggett2017}, thereby complicating the construction of precise and accurate CMDs.

Likewise, while synthetic photometry is generally very reliable for brighter stars and warmer brown dwarfs, direct photometry extracted from imaging observations and synthetic photometry obtained from integrated-light spectra do not always agree for Y dwarfs.  Discrepancies on the order of several tenths of a magnitude have been observed in some bands, both with \textit{HST} \citep{Schneider2015} and \textit{JWST} \citep{Beiler2024,Albert2025}, attributed to the unusual spectral shapes of Y dwarfs. For such cold objects, combining heterogeneous methods of photometric measurements therefore inevitably introduces additional uncertainties that propagate into CMDs, ultimately limiting the precision and accuracy with which intrinsic population‑level trends can be identified.

Eliminating such potential sources of observational systematics is essential for isolating and interpreting the true astrophysical scatter within the underlying Y-dwarf sequences. Our homogeneous \textit{HST} data were acquired entirely with a single instrument and filter system, and processed through uniform extraction and calibration methods, enabling a sizeable sample of 21 Y dwarfs to be placed on common CMDs with consistent internal precision and accuracy.

\subsection{Observed trends in the Y-dwarf sequence}

A notable feature of the NIR sequence is the blueward progression in F105W--F125W ($Y$--$J$) with decreasing temperature \citep{Liu2012,Lodieu2013,Schneider2015,Schneider2016,LuhmanEsplin2016,Leggett2015,Leggett2021}, attributed to the condensation of potassium in the late-T and Y-dwarf temperature regime \citep{Burningham2010,Leggett2010,Phillips2020}. This condensation weakens the potassium resonance doublet at $\sim$0.77\,$\mu$m, lowering the opacity of the broad red-wing, which brightens the $Y$ band. A reversal of the blueward $Y$--$J$ trend occurs at colder temperatures, with colours becoming redder toward the regime of WISE\,0855$-$0714 with the collapse of the Wien tail at T$_\mathrm{eff} < 350$\,K \citep{Burrows2003}. Our data highlight this sequence very clearly, and align with earlier hints \citep{Schneider2016} that this reversal might begin to emerge at the cold end of our core sample.
Likewise, our F125W--F160W colours reproduce the established $J$--$H$ behaviour. Following the plateau seen for mid‑ to late‑T dwarfs in these colours, Y dwarfs transition into a monotonic redward trend \citep{Cushing2011,Cushing2014,DupuyLiu2012,Kirkpatrick2012,Leggett2013,Leggett2017,Schneider2016,LuhmanEsplin2016}, observed with improved tightness in our \textit{HST} data.

The F125W--[4.5] vs. $M_\mathrm{F125M}$ CMD traces a tight monotonic sequence that is well correlated with effective temperature thanks to the long wavelength baseline that combined NIR--MIR colours provides \citep{Martin2018,Kirkpatrick2019,Leggett2021}, extending the shift to redder $J$--[4.5] colours with decreasing absolute magnitude seen in earlier spectral types \citep{Kirkpatrick2011,Kirkpatrick2012,Leggett2013}. All atmospheric grids in Figures~\ref{f:CMD_models_ATMO} to \ref{f:CMD_models_ELFOWL}, equilibrium and non‑equilibrium models alike, converge on this monotonic redward progression and reproduce observed colours successfully, although some model assumptions, in particular metallicity, introduce variations in the position of the slope (see also \citealp{Tinney2014,Schneider2015,Leggett2017}).
On the other hand, the MIR [3.6]--[4.5] vs. $M_\mathrm{F125W}$ diagram shows more substantial scatter and is less strongly correlated, as shown in Section~\ref{MIR_CMDs}. This effect is attributed to complex flux redistribution at these wavelengths arising from vertical mixing, and the strong sensitivity of the 3.6\,$\mu$m CH$_4$ absorption band to disequilibrium chemistry, gravity and metallicity, although the [3.6] fluxes remain consistently too faint in current models for brown dwarfs cooler than $\sim$600\,K \citep{Tremblin2015,Leggett2019,Phillips2020,Leggett2021,Leggett2023}. 

A handful of noticeable outliers to the core population were identified, in particular WISE\,0535$-$7500 and WISE\,1828$+$2650. Their positions redward of the main locus in NIR--MIR CMDs ---or equivalently appearing overluminous for their colours--- together with their high effective temperatures (inferred from \textit{JWST} spectra extending into the MIR) that seem inconsistent with their locations along the NIR sequences, reinforce earlier suspicions that these systems might be unresolved binaries \citep{Beichman2013,Leggett2013,Tinney2014,Leggett2017,Kirkpatrick2019,Cushing2021}, akin to the 1-au WISE\,0336$-$0143\,AB binary system resolved with \textit{JWST} \citep{Calissendorff2023} that occupies a similar overluminous region in \textit{HST}-\textit{Spitzer} CMDs.  Since these objects do not prominently stand out in the NIR sequences, they are more likely to be hosting colder secondaries with small flux contributions in the NIR, that become more significant in the MIR, thereby boosting the combined 4.5\,$\mu$m fluxes of the systems.

The significant luminosity gap between the bulk of the Y‑dwarf population studied here and the extreme WISE\,0855$-$0714 highlights the need for further discoveries and detailed characterisation of the coldest Y dwarfs (e.g., \citealp{Calissendorff2023,Leggett2023}), in order to bridge this regime and clarify the atmospheric processes that shape the cooler end of the sequence. While the reddest and faintest Y dwarfs remain challenging for \textit{HST} (e.g., the F105W and F125W non‑detections of WISE\,0830$+$2837; \citealp{BardalezGagliuffi2020}), several recent discoveries populating the regime below $\sim$350\,K \citep{Meisner2020,Meisner2020b,Kirkpatrick2021} have now been also observed with \textit{HST}/WFC3, although primarily in the F110W filter (GO‑16243, PI Marocco), for which we lack comparable photometry across most of our sample. Securing homogeneous sets of observations with the same instruments and filters for the entire Y‑dwarf population will be needed to establish a consistent baseline, reduce systematic offsets, and enable robust comparisons across the full sequence.

\subsection{The effects of metallicity and C/O ratio}

Across all model families tested, low metallicity grids consistently yield the best overall fits to the global Y‑dwarf population. Yet it is evident that no single metallicity reproduces all NIR and MIR colour indices simultaneously: sub-solar metallicities are needed to match the observed blue $Y$--$J$ and [3.6]--[4.5] colours, while solar and super-solar abundances best reproduce $Y$--$H$ and $J$--$H$ colours. This suggests that low metallicities tend to improve agreement in the spectral regions where the largest discrepancies arise, and that the outcome of the fits is likely sensitive to the particular flux bands included in the analysis. We therefore interpret these results not as direct evidence that Y‑dwarf atmospheres are intrinsically all metal‑poor, but rather that certain low‑metallicity prescriptions mimic missing or incomplete physics in specific wavelength ranges. In other words, the apparent preference for low metallicity may reflect compensating effects in the models, rather than a true population‑level abundance trend.

For instance, \citet{Morley2018} found that low‑metallicity models with half‑solar C/O ratios provided the best matches to the mid‑infrared SED of WISE\,0855$-$0714. Both effects act to deplete the methane abundance, thereby weakening the CH$_4$ absorption feature around 3--4\,$\mu$m and brightening the [3.6] region, although similar effects can be obtained by heating the upper atmosphere through vertical mixing (see below). \citet{Cushing2021} reached similar conclusions in the atmospheric modelling of the suspected binary WISE\,1828$+$2650, finding that models with sub-solar metallicities and C/O ratios better reproduce the broad MIR [3.6] and [4.5] photometry.
In a similar way, the apparent preference for low‑metallicity grids in our fits likely reflects their ability to better reproduce the MIR \textit{Spitzer} photometry. At the same time, metal depletion would also reduce potassium opacity (see below) and enhance the emerging flux in the $Y$ band. Taken together, these effects would yield closer agreement with the observations in both the F105W and [3.6] filters, though through distinct physical implications arising from lower metallicities. It is thus possible that separate independent processes (e.g., the treatment of potassium chemistry and vertical mixing in the upper atmosphere) could induce those same effects on the NIR and MIR spectral regions, and successfully explain observations without the need for a strong metal paucity across the entire Y-dwarf population.

Furthermore, the interplay between different model parameters becomes evident in the \texttt{Sonora Elf\,Owl fits}, where decreasing metallicity is accompanied by a shift towards higher C/O ratios (Figure~\ref{f:chi2-map}). At $\log(Z/Z_\odot)=-$0.5, the best fits favour solar C/O values (0.458), whereas at $\log(Z/Z_\odot)=-$1.0 the models prefer super‑solar C/O ratios (0.687) for comparably good population‑level fits. Based on the way C/O ratio is varied in the \texttt{Sonora} models \citep{Marley2021}, this behaviour suggests that when the metallicity is too low, the carbon abundance must be boosted to recover agreement with the observed photometry. We note that while we only performed fits at fixed grid values for these model parameters, it is possible that intermediate C/O ratios and metallicities could provide even better fits to the global observed population. Nonetheless, this emphasises the complex degeneracies between model settings in reproducing Y‑dwarf colours, and cautions that combinations of currently explored model settings that appear as better fits may be highlighting compensating effects rather than true underlying atmospheric properties.

\subsection{The effects of disequilibrium chemistry and vertical mixing}

Disequilibrium chemistry driven by vertical mixing is now widely recognised to play a central role in shaping Y‑dwarf spectra. Under equilibrium conditions, CH$_4$ and NH$_3$ should dominate at the cold effective temperatures of Y dwarfs, with depleted CO and N$_2$. Instead, observations reveal significant absorption from carbon monoxide in the 3--5\,$\mu$m region and suppressed methane and ammonia features, which is seen as evidence that vigorous vertical mixing transports long-lived species upward from deeper, hotter layers of the atmosphere \citep{Phillips2020,Leggett2021,Beiler2024a,Kothari2024}. The enhanced CO absorption suppresses flux in the [4.5] band, while the quenching of CH$_4$ and NH$_3$ reduces opacity in the $H$ and [3.6] bands, making them brighter and fundamentally altering colours at these wavelengths, adding an extra layer of complexity to the processes mentioned above. \citet{Miles2020} found that the coolest brown dwarfs below 400\,K have higher mixing strengths, while warmer brown dwarfs show weaker atmospheric mixing, suggesting that disequilibrium chemistry becomes increasingly important towards the coldest end of the Y‑dwarf sequence.

It is thus somewhat surprising that equilibrium chemistry grids provided the best fits within the \texttt{ATMO\,2020} and \texttt{Lacy\,\&\,Burrows} models. We attribute this effect to the closer match that models in chemical equilibrium appear to provide to the F105W--F160W ($Y$--$H$) colours. Grids in disequilibrium tend to predict $Y$--$H$ colours that are redder than observed, which is likely a result of the brightening of the $H$-band under non-equilibrium conditions. Among the discrete grids shown in Figure~\ref{f:CMD_models_ATMO} to \ref{f:CMD_models_ELFOWL}, the only non-equilibrium models that seem to reproduce the observed blue $Y$--$H$ colours are the super-solar metallicity \texttt{Elf\,Owl} spectra from the \texttt{Sonora} models, though these higher metallicity grids are highly disfavoured in the overall fits across all considered photometric bands. It is also interesting to note that the best fits results within the \texttt{Elf\,Owl} grids, at sub-solar metallicities, show a preference towards lower vertical mixing strengths, which primarily impact the MIR [3.6]--[4.5] colours but exert limited influence on the NIR colours inspected here. Again this points to the idea that various atmospheric settings in the models appear to compensate for missing or incomplete physics in different spectral regions, rather than reflecting true population‑level properties.

\subsection{Potassium in the Y band}

There is a consistent discrepancy between all the model families and the observed Y-dwarf population in CMDs involving the F105W filter, more commonly referred to as the $Y$ band. While most models predict a blueward trend in F105W--F125W with lower temperature comparable to the observed sequence, they predict F105W--F125W colours (and in most chemical disequilibrium cases, F105W--F160W colours too) that are too red compared to the observations \citep{Schneider2015,Phillips2020,Leggett2021,Leggett2025}. The flux in the $Y$ band is shaped by the far-red wing of the potassium resonance doublet located at 0.77\,$\mu$m. The line shapes of this doublet are determined by the potential field of $\mathrm{H_2}$ perturbing the ground and excited states of the alkali atom. In the cool, dense atmospheres of late-T and Y dwarfs, these resonance lines become broadened out to thousands of angstroms away from the line core, shaping the red-optical and near-infrared flux by controlling the level of flux in the $Y$ band. As such, Lorentzian line profiles are not sufficient in modelling the collisional broadening effects on alkali metals like potassium, and more detailed quantum mechanical calculations of the interaction potentials of these collisions are required to accurately model the potassium resonance line shapes \citep{Burrows2003, Allard2007, Allard2016, Allard2019}. Different line profile treatments are known to have a large impact and therefore uncertainty on the predicted emission spectra of brown dwarfs \citep{Baudino2017, Phillips2020}. All the models used in this work use the latest potassium line profiles from \citet{Allard2016}. Given the models predict F105W--F125W and F105W--F160W colours that are too red, this could indicate that the potassium line profiles of \citet{Allard2016} are too strong, with too much opacity in the $Y$ band. 

Alternatively, the issue may lie in the abundance of potassium rather than the opacity. In pure chemical equilibrium, K remains in the gas phase until local atmospheric temperatures of $\sim$1300\,K, when it condenses into $\mathrm{KAlSi_3O_8}$. However, in rainout equilibrium, the formation of magnesium silicate and corundum condensates deplete Al and Si from the atmosphere, preventing the formation of $\mathrm{KAlSi_3O_8}$, leaving potassium in the gas phase until it condenses into KCl at $\sim$700\,K \citep{Burrows2001}. Evidence for rainout chemistry of alkali metals has been found by retrieval studies of Y dwarfs \citep{Zalesky2019, Zalesky2022}, and all the models used in this work include rainout chemical equilibrium. Despite this, the models still struggle to reproduce the $Y$--$J$ colours of Y dwarfs. It was noted by \citet{Phillips2020} and \citet{Leggett2021} that reducing the potassium abundance by approximately an order of magnitude can help rectify the model-data discrepancy in the $Y$ band. Therefore, these results could indicate that the current modelling of the equilibrium abundance and/or condensation of potassium is slightly incorrect.

\section{Conclusions}
\label{conclusions}

In this third paper of our series, we presented a homogeneous set of high‑precision \textit{HST}/WFC3 near‑infrared photometry for 21 Y dwarfs, providing one of the largest uniform photometric dataset across the F105W ($Y$), F125W ($J$), and F160W ($H$) filters for these objects, in addition to some F110W, F127M and F140W flux measurements for a subset of the sample. Combined with robust new parallaxes from \citetalias{Fontanive2025}, these measurements reveal a well‑defined Y‑dwarf locus in colour–magnitude space, observed with minimal scatter compared to previous observations. We recover the blueward trend in $Y$--$J$ with decreasing temperature, with possible hints of a reversal emerging around T$_\mathrm{eff}$$\sim$350\,K, while $J$--$H$ continues to redden monotonically across the cooling sequence. The improved tightness of our observed sequences across NIR bands reflects the benefits of the highly precise, internally consistent photometry achieved here with \textit{HST}.

Population-level comparisons to atmospheric grids from the \texttt{ATMO} \citep{Phillips2020,Leggett2021}, \texttt{Lacy\,\&\,Burrows} \citep{LacyBurrows2023} and \texttt{Sonora Elf\,Owl} \citep{Mukherjee2024} models show that no single combination of parameters reproduces all observed colours simultaneously. In particular, models predict F105W--F125W ($Y$--$J$) and F105W--F160W ($Y$--$H$) colours that are almost always too red, a discrepancy attributed to shortcomings in the treatment of potassium opacity or chemistry affecting the blue wing of the $Y$ band. Apparent preferences for low metallicity or equilibrium chemistry across the full population likely reflect compensating effects in specific bands (e.g., $Y$--$H$ or MIR fluxes) rather than true population‑level properties, pointing to degeneracies between metallicity, C/O ratio, cloud physics, and mixing strength. While some diversity in atmospheric properties is naturally expected across the Y-dwarf population, the systematic offsets between observed loci and model predictions highlight that the dominant physical and chemical mechanisms driving the global photometric properties of Y-dwarfs remain incompletely captured in current models.

These results provide a robust empirical baseline to improve atmospheric models for the Y‑dwarf population, although a significant gap in temperature and photometric characteristics remains to be filled between the bulk of our studied sample and the coldest known object, WISE\,0855$-$0714. Our results highlight that uniform sets of high-precision observations are needed to constrain the evolution of spectro-photometric properties in Y dwarfs along the cooling sequence with high levels of confidence, which is in turn required to isolate the effects of temperature from secondary physical and atmospheric parameters. The expanding collections of \textit{JWST} imaging and spectroscopy are enabling increasingly comprehensive tests of atmospheric chemistry and dynamics in individual objects at the cold end of the Y-dwarf population \citep{Leggett2023,Beiler2024a,Kothari2024,Lew2024,Luhman2024,Lueber2026}. Analysing these datasets at the population level will remain equally important for a complementary perspective into bulk trends across the full Y-dwarf sequence, and to empirically improve theoretical models at the lowest effective temperatures, thereby providing crucial context for interpreting extra‑solar giant planets in the Y‑dwarf temperature regime \citep{Matthews2024,BardalezGagliuffi2025,Lagrange2025}.

\section*{Data Availability}

All observational data used in this work are publicly available. 
Reduced stacked images produced for each epoch/filter combination for
each target are provided at \url{https://github.com/cfontanive/BDs_HST}.

\section*{Acknowledgements}

CF gratefully acknowledges support from the School of Physics \& Astronomy, University of Edinburgh, through the Elizabeth Gardner Fellowship, as well as the Trottier Family Foundation and the Trottier Institute for Research on Exoplanets that supported her during the initial stage of this work through her Trottier Postdoctoral Fellowship.
The authors acknowledge the support from STScI award \#HST-GO-16229.002-A.
This research is based on observations made with the NASA/ESA \textit{Hubble Space Telescope} obtained from the Space Telescope Science Institute, which is operated by the Association of Universities for Research in Astronomy, Inc., under NASA contract NAS 5–26555. These observations are associated with programs GO-17466, 17403, 17080, 16229, 15201, 14157, 14233, 13802, 13428, 13178, 12972, 12970, 12873, 12815, 12544, 12330. Some of the data presented in this paper were obtained from the Mikulski Archive for Space Telescopes (MAST).
This work has made use of data from the European Space Agency (ESA) mission {\it Gaia} (\url{https://www.cosmos.esa.int/gaia}), processed by the {\it Gaia} Data Processing and Analysis Consortium (DPAC, \url{https://www.cosmos.esa.int/web/gaia/dpac/consortium}). Funding for the DPAC has been provided by national institutions, in particular the institutions participating in the {\it Gaia} Multilateral Agreement.
This research has benefited from the Y Dwarf Compendium maintained by Michael Cushing at \url{https://sites.google.com/view/ydwarfcompendium/}.



\bibliographystyle{mnras}
\input{biblio.bbl}










\bsp	
\label{lastpage}
\end{document}